\documentclass[12pt]{JHEP3}
\usepackage{amssymb,graphics,epsfig,rotating,subfigure,sidecap}
\def\beq   {\begin{equation}}
\def\eeq   {\end{equation}}
\def\beqd  {\begin{displaymath}}
\def\eeqd  {\end{displaymath}}
\def\beqaa {\begin{eqnarray}}
\def\eeqaa {\end{eqnarray}}

\def\ti  {\tilde}

\def\sf  {\ti f}
\def\sq  {\ti q}

\def\sl  {\ti \ell}

\def\t   {\theta}

\def\orp {\overrightarrow{p}}

\def\sz{\ifmmode{\tilde{\chi}^0} \else{$\tilde{\chi}^0$} \fi}
\def\sw{\ifmmode{\tilde{\chi}} \else{$\tilde{\chi}$} \fi}

\newcommand{\be}[1]{\begin{equation} \label{(#1)}}
\newcommand{\ee}{\end{equation}}
\newcommand{\baq}[1]{\begin{eqnarray} \label{(#1)}}
\newcommand{\eaq}{\end{eqnarray}}

\newcommand{\ba}{\begin{array}}
\newcommand{\ea}{\end{array}}

\title{Probing CP Violation with and without\\ Momentum Reconstruction at the LHC}

\author{G.~Moortgat-Pick\\ IPPP, University of Durham, Durham DH1 3LE, UK\\ E-mail: \email{g.a.moortgat-pick@durham.ac.uk}}
\author{K.~Rolbiecki\\ IPPP, University of Durham, Durham DH1 3LE, UK\\ E-mail: \email{krzysztof.rolbiecki@durham.ac.uk}}
\author{J.~Tattersall\\ IPPP, University of Durham, Durham DH1 3LE, UK\\ E-mail: \email{jamie.tattersall@durham.ac.uk}}
\author{P.~Wienemann\\ Department of Physics, University of Bonn, Nussallee 12, D-53115 Bonn, Germany\\ E-mail: \email{wienemann@physik.uni-bonn.de}}

\abstract{We study the potential to observe CP-violating effects in SUSY cascade decay chains at the LHC. We consider squark and gluino production followed by subsequent decays into neutralinos with a three-body leptonic decay in the final step. Asymmetries composed by triple products of momenta of the final state particles are sensitive to CP-violating effects. Due to large boosts these asymmetries can be difficult to observe at a hadron collider. We show that using all available kinematic information one can reconstruct the decay chains on an event-by-event basis even in the case of 3-body decays, neutrinos and LSPs in the final state. We also discuss the most important experimental effects like major backgrounds and momentum smearing due to finite detector resolution. We show that with 300~fb$^{-1}$ of collected data, CP violation may be discovered at the LHC for a wide range of the phase of the bino mass parameter $M_1$.}


\preprint{arXiv:0908.2631 \\ DCPT-09-108 \\ IPPP-09-54}
\begin{document}


\section{Introduction}\label{sec:introduction}

Supersymmetry (SUSY)~\cite{Golfand:1971iw,Wess:1973kz,Wess:1974tw} is one of the best motivated extensions to the Standard Model (SM). According to the latest fits~\cite{Buchmueller:2008qe,Buchmueller:2009fn,Bechtle:2009ty} to the precision observables a rather light scale is indicated and is in reach of the Large Hadron Collider (LHC). If SUSY happens to be realised in nature, the task will be to pin down the specific characteristics of the underlying model which will require many detailed studies.
One of the interesting issues in this context is CP violation. While the observed amount of CP violation in the $K$ and $B$ sectors can be accommodated within the SM, another piece of evidence, the baryon asymmetry of the universe, requires a new source of CP violation~\cite{Cohen:1993nk,Gavela:1994dt,Rubakov:1996vz}.

The Minimal Supersymmetric Standard Model (MSSM)~\cite{MSSM,Haber:1984rc,Martin:1997ns} has about one hundred free parameters and a large number of these may have non-zero CP-violating phases, see e.g.\ \cite{Ibrahim:2007fb}. Many of the phases are unphysical in the sense that they can be rotated away by a redefinition of the fields but not all can be systematically removed in this way. For example in the sectors of the charginos and neutralinos, the supersymmetric partners of Higgs bosons and $SU(2)\times U(1)$ gauge bosons, we can have up to three new phases. These include the phases of the $U(1)$ gaugino mass parameter $M_1$, $SU(2)$ gaugino mass parameter $M_2$ and the Higgsino mass parameter $\mu$. One of them can be rotated away and therefore we choose the convention where $M_2$ is kept real. Hence we have
\begin{equation}
M_1 = |M_1| e^{i \phi_1}\,, \qquad \mu = |\mu| e^{i \phi_\mu} \,.
\end{equation}
The other possible sources of complex parameters in the MSSM Lagrangian are trilinear couplings $A_f$ but they are not relevant to this study~\cite{Ellis:2008hq}.

Certain combinations of the CP-violating phases are constrained by the experimental upper bounds on various electric dipole moments (EDMs), see e.g.~\cite{Ellis:2008zy}. Ignoring possible cancellations, the most severely constrained phase is that of $\mu$ which contributes to the EDMs at the one-loop level and has a $\mathrm{tan}\beta$ enhancement. In general for $\mathcal{O}(100)$~GeV masses, $|\phi_{\mu}|\lesssim0.01\pi$ and we therefore set $\phi_{\mu}=0$ throughout our study. The phase of $M_1$ has weaker constraints. Large phases can be accommodated in models with heavy first and second generation squarks and sleptons ($>$TeV) or if accidental cancellations occur~\cite{Kizukuri:1992nj,Ibrahim:1998je,Ibrahim:1999af,Brhlik:1998zn,Abel:2001vy,Arnowitt:2001pm}. Here we study the complete range of CP phases in order to see the general dependencies exhibited by our observables and the luminosity required to observe these within the LHC environment. Nevertheless, we want to stress that in the chosen scenario experimental bounds from EDMs can be evaded by arranging cancellations between various supersymmetric contributions for any value of $\phi_1$~\cite{Lee:2003nta,Lee:2007gn,Ellis:2008zy}. 

CP-odd observables are the unambiguous way of discovering hints of complex parameters in the model. One class of these observables include rate asymmetries of cross sections, branching ratios and angular distributions. The other possibility are the observables that can be defined using triple products of momenta and/or spins of particles, see~\cite{Kittel:2009fg,Hesselbach:2007dq,Kraml:2007pr} for a recent review. In any case studies of CP-violating phases will be challenging at the LHC~\cite{Langacker:2007ur,Ellis:2008hq,Deppisch:2009nj} and a precise determination of them is expected to be only possible at a future $e^+e^-$ linear collider. 

An interesting example of using triple product correlations is studying the 3-body decay of a neutralino $\tilde{\chi}_2^0$~\cite{AguilarSaavedra:2004hu,AguilarSaavedra:2003ur,Choi:2005gt,Bartl:2004jj}. Since the asymmetry is maximal in the rest frame of the decaying neutralino, the central idea of~\cite{AguilarSaavedra:2004hu,AguilarSaavedra:2003ur} was the full kinematic reconstruction of the production and decay process. Hence the CP-sensitive observables are calculated~\cite{AguilarSaavedra:2004hu,AguilarSaavedra:2003ur} in the rest frame of the decaying neutralino at an $e^+e^-$ collider. This can be done due to the clean experimental environment and the well-known initial state at such a machine. As this is not the case at the LHC, these observables are much more challenging due to the proton structure and further experimental uncertainties, hence making the observation of CP-violating effects very difficult. However, in the present analysis we show how a similar approach can be applicable at hadron colliders. By applying full event reconstruction, the observation of CP-odd asymmetries becomes feasible.

In many scenarios the highest cross section at the LHC is predicted for the associated production of gluinos and squarks of the first two generations. If squarks can decay to a neutralino $\tilde{\chi}_2^0$ in the process
\begin{equation}
 \tilde{q}_i \to \tilde{\chi}^0_2 + q
\end{equation}
one can expect an abundance of neutralinos. Large statistics will help to overcome the limitations of a similar analysis of the stop carried out in \cite{Ellis:2008hq}.
The production of the neutralino $\tilde{\chi}^0_2$ is followed by the subsequent three-body decay
\begin{equation}
 \tilde{\chi}^0_2 \to \tilde{\chi}^0_1 \ell^+\ell^-.
\end{equation}
The 3-body decay is discussed since sizeable contributions to CP violation are expected from interference diagrams and it offers the opportunity to use CP-sensitive observables based on triple product correlations. The process is sensitive to any phase that enters the neutralino sector ($\phi_{1}$ and $\phi_{\mu}$), but we set $\phi_{\mu}$ to zero as discussed above and concentrate on the effect of $\phi_{1}$. To extract information about the phase of $M_1$ we analyse the following triple product of momenta from the decay products of the $\tilde{q}_i$,
\begin{equation}
 \mathcal{T}= \vec{p}_q \cdot (\vec{p}_{\ell^+} \times \vec{p}_{\ell^-}).
	\label{eq:TripProd}
\end{equation}
It originates from the covariant product present in the amplitude and can be expanded in terms of explicit momenta,
\begin{equation}
 \epsilon_{\mu\nu\rho\sigma}p_a^\mu p_b^\nu p_c^\rho p_d^\sigma \longrightarrow  E_a\;\overrightarrow{p_b}\cdot(\overrightarrow{p_c}\times\overrightarrow{p_d}) \pm \dots
 \label{eq:EpsExpan}
\end{equation}
When CP phases appear a non-zero asymmetry can be constructed from final state momenta using the triple product in Eq.~(\ref{eq:TripProd}) as explained below.

The triple product $\mathcal{T}$ changes sign under the na\"{\i}ve\footnote{Na\"{\i}ve time reversal $\mathrm{T}_N$ reverses the momenta and spins of all particles without interchanging the initial and final states, see e.g.~\cite{Atwood:2000tu}. Recall that under the true time reversal also the initial and final states are interchanged.} time reversal $\mathrm{T}_N$ and consequently it is a $\mathrm{T}_N$-odd observable. Neglecting possible absorptive phases due to loops, Breit-Wigner propagators and final state interactions, that are expected to be small in the leptonic neutralino decay, $\mathrm{CPT}_N$ invariance is equivalent to CPT invariance~\cite{Atwood:2000tu} and hence a measurement of a $\mathrm{T}_N$-odd observable implies CP violation. In addition as these are CP-odd quantities any measurement of the phase made by this method not only reveals the magnitude of the phase but also the sign. It is important to note that the triple product correlations are a tree-level effect and therefore large asymmetries can be expected~\cite{Bartl:2004jj,Bartl:2002hi,Bartl:2003ck,Bartl:2006yv}.

To be able to successfully observe CP violation we need to isolate squarks from antisquarks as the CP-conjugated process will have an asymmetry of the opposite sign. Normally this is done via charge identification but this will be impossible using the above decay chain with the light quark in the final state. However, for the first two generations, squark production typically dominates over anti-squark production due to the valence quarks present in the proton. Consequently our sample will have a majority of squarks (around 80\% in the analysed scenario) and the asymmetry will not be washed out. The proportion depends on the masses of squarks but this does not affect our analysis significantly.

One of the major problems when trying to measure CP-sensitive observables with triple product correlations at the LHC are the large undetermined boosts given to produced particles. CP asymmetries have the largest value in the rest frame of the intermediate particle and the effect of the boost produces a severe dilution factor. To remove this dilution factor we have to boost all the particles in the triple product ($q, \ell^+, \ell^-$) into the rest frame of the intermediate particle in the decay ($\tilde{\chi}^0_2$ in our case). In SUSY decay chains this is complicated due to the presence of two LSPs and possible neutrinos that escape detection. Once the missing momenta are reconstructed it is then trivial to find the momentum of the particle that we are interested in and boost back into its rest frame. In principle this should return the CP asymmetry to the magnitude one would expect if the initial particle was produced at rest.

So far in the literature the momentum reconstruction was used to determine the unknown masses of SUSY particles in the case when the decay chain proceeded via two-body decays~\cite{Nojiri:2003tu,Nojiri:2007pq,Cheng:2008mg,Cheng:2009fw}. In this analysis we change the approach and assume that the masses are already known with a reasonable precision from other measurements at the LHC. The new complications are three-body decays of supersymmetric particles (charginos and neutralinos) and possible missing neutrinos which means that we have less on-shell constraints. However, as we show in this paper, even in such a case momentum reconstruction is still possible after the inclusion of experimental uncertainties. Having all the momenta reconstructed one can perform a more detailed analysis of particle properties which could open a window for a more accurate determination of the underlying model at the LHC.

Finally, in order to assess the feasibility of completing this study in the demanding experimental environment of the LHC we perform a simulation of our process using the Monte Carlo event generator \texttt{Herwig++}~\cite{Bahr:2008pv,Bahr:2008tx}. We apply simple selection cuts and smear final state particle momenta to account for finite detector resolution. Moreover, we make an analysis of the impact of the most important background processes relevant for our study. All of the analysis was done assuming a centre-of-mass energy of 14~TeV but it would not change markedly if we moved to 10~TeV except for an increase in the luminosity we require to see a statistically significant result.

The paper is organised as follows. We begin by describing the process under consideration in Section~\ref{sec:formalism} and  discuss the triple product correlations in more detail. In Section~\ref{sec:MomRec} we present the method we use to reconstruct the momenta of the invisible particles in our decay chain. Section~\ref{sec:results} discusses the potential for a measurement at the LHC both with and without the momentum reconstruction technique. Possible experimental factors that need to be considered and our final results are explained in Section~\ref{sec:experimental}. Finally, the Appendices list the Lagrangian, resulting couplings and give explicit matrix elements including the full spin correlations that are required in our study.


\section{Formalism}
\label{sec:formalism}

\subsection{The process studied and the amplitude squared }
\label{sec:process}

\begin{figure}[t!]
\begin{center}
\hspace{-0.4cm}
\begin{minipage}[t]{5cm}
\begin{center}
{\setlength{\unitlength}{0.8cm}
\begin{picture}(5,6)
\put(0,2){\epsfig{file=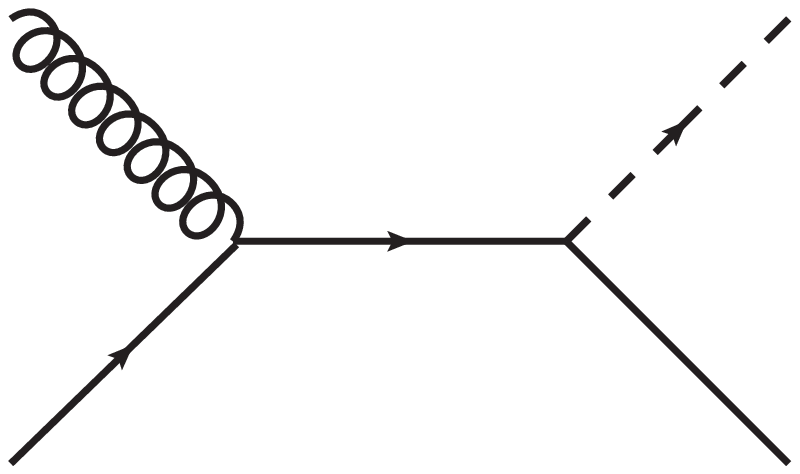,scale=0.5}}
\put(5.3,1.9){{\small $\tilde{g}$}}
\put(-0.1,1.9){{\small $q$}}
\put(5.3,5.0){{\small $\tilde{q}_L$}}
\put(-0.1,5.0){{\small $g$}}
\put(2.6,3.8){{\small $q$}}
\end{picture}}
\end{center}
\end{minipage}
\hspace{0.cm}
\begin{minipage}[t]{5cm}
\begin{center}
{\setlength{\unitlength}{0.8cm}
\begin{picture}(5,6)
\put(0,2){\epsfig{file=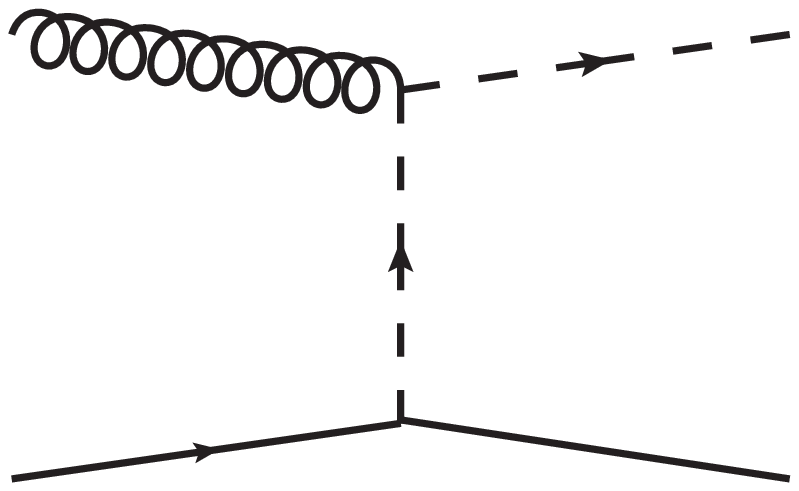,scale=0.5}}
\put(5.3,1.9){{\small $\tilde{g}$}}
\put(-0.1,1.9){{\small $q$}}
\put(5.3,5.0){{\small $\tilde{q}_L$}}
\put(-0.1,5.0){{\small $g$}}
\put(2.9,3.4){{\small $\tilde{q}_L$}}
\end{picture}}
\end{center}
\end{minipage}
\begin{minipage}[t]{5cm}
\begin{center}
{\setlength{\unitlength}{0.8cm}
\begin{picture}(5,6)
\put(0,2){\epsfig{file=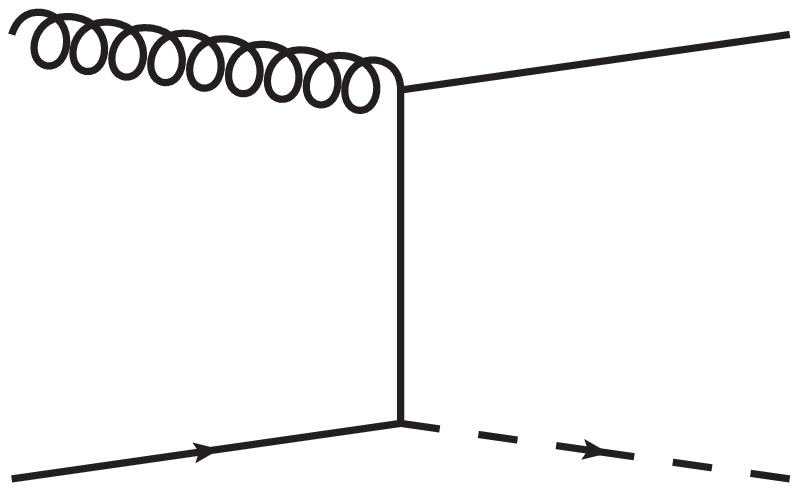,scale=0.5}}
\put(5.3,1.9){{\small $\tilde{q}_L$}}
\put(-0.1,1.9){{\small $q$}}
\put(5.3,5.0){{\small $\tilde{g}$}}
\put(-0.1,5.0){{\small $g$}}
\put(2.9,3.4){{\small $\tilde{g}$}}
\end{picture}}
\end{center}
\end{minipage}
\end{center}
\vspace{-1.7cm}
\caption{\label{Fig:FeynProdA}Feynman diagrams for the associated production process of squarks and gluinos at the LHC, $gq \to \tilde{g} \tilde{q}_L$.}
\end{figure}

One of the dominant SUSY channels at the LHC is associated squark-gluino production. At the tree level the production process Eq.~(\ref{eq-prod}) proceeds via the light quark exchange in the direct channel and squark/gluino exchange in the $t$ channel, as seen in Fig.~\ref{Fig:FeynProdA}, 
\begin{equation}
gq \to \tilde{g} \tilde{q}_L.
\label{eq-prod}
\end{equation}
We study the case where the squark subsequently decays via the following chain:
\begin{equation}
\tilde{q}_L \to \tilde{\chi}^0_j + q \to \tilde{\chi}^0_1 \ell^+\ell^- + q.
\label{eq-decays} 
\end{equation}

\noindent The first step in the cascade decay chain is the two-body process $\tilde{q}_L \to q
\tilde{\chi}^0_2$. Here, CP-violating couplings of the $\tilde{\chi}^0_2$ enter and are dominated by the phase $\phi_{1}$, see Appendix~\ref{sect:squarkdecay}. In addition, the spin vector of the $\tilde{\chi}^0_2$ has to be explicitly included in the amplitude since the full spin correlations have to be taken into account.

\begin{figure}[t!]
\begin{center}
\hspace{-0.5cm}
\begin{minipage}[t]{5cm}
\begin{center}
{\setlength{\unitlength}{0.8cm}
\begin{picture}(5,8)
\put(-5.2,-5.2){\includegraphics{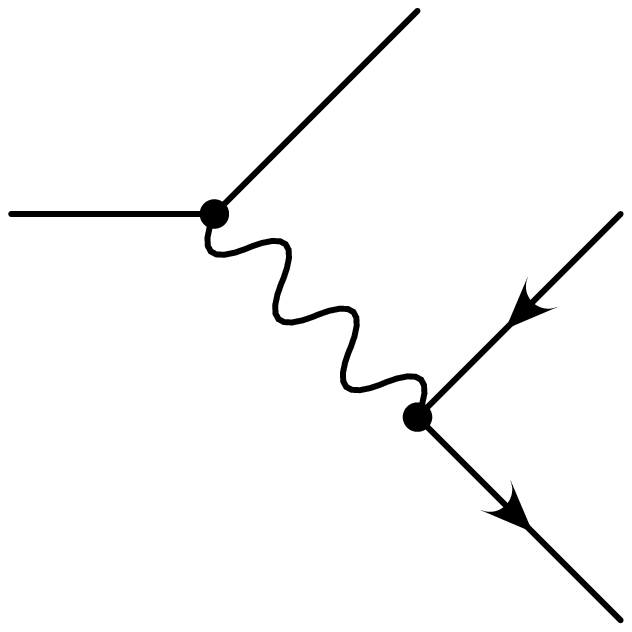}}
\put(4.7,1.7){{\small $\ell^{-}$}}
\put(0.3,5.5){{\small $\tilde{\chi}^0_{2}$}}
\put(4.7,5.5){{\small $\ell^{+}$}}
\put(3.7,7.1){{\small $\tilde{\chi}^0_{1}$}}
\put(2.2,3.8){{\small $Z^0$}}
\end{picture}}
\end{center}
\end{minipage}
\hspace{-0.5cm}
\begin{minipage}[t]{5cm}
\begin{center}
{\setlength{\unitlength}{0.8cm}
\begin{picture}(5,8)
\put(-5.2,-6){\includegraphics{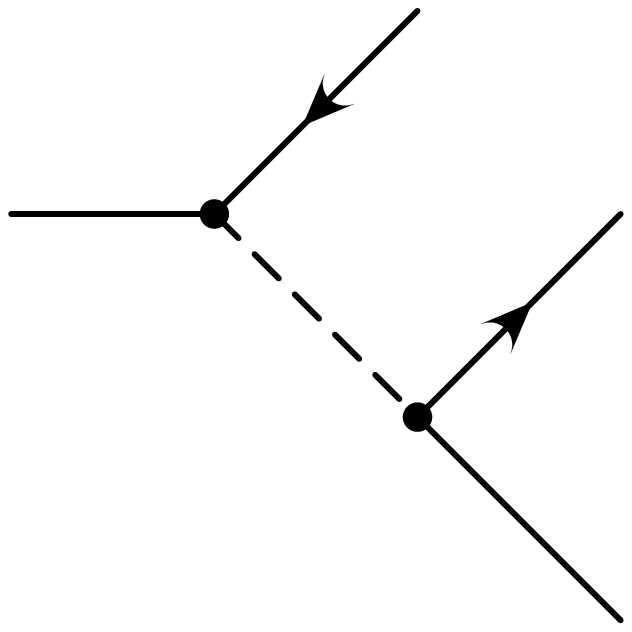}}
\put(4.7,5.5){{\small $\ell^{-}$}}
\put(3.7,7.1){{\small $\ell^{+}$}}
\put(4.7,1.5){{\small $\tilde{\chi}^0_{1}$}}
\put(0.3,5.6){{\small $\tilde{\chi}^0_{2}$}}
\put(2.1,3.9){{\small $\tilde{\ell}_{L,R}$}}
\end{picture}}
\end{center}
\end{minipage}
\begin{minipage}[t]{5cm}
\begin{center}
{\setlength{\unitlength}{0.8cm}
\begin{picture}(5,8)
\put(-5.2,-6){\includegraphics{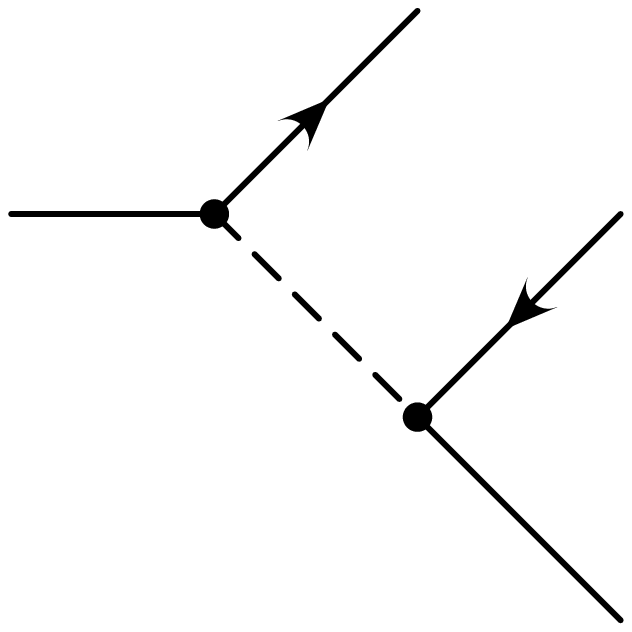}}
\put(3.7,7.1){{\small $\ell^{-}$}}
\put(4.7,5.5){{\small $\ell^{+}$}}
\put(0.3,5.6){{\small $\tilde{\chi}^0_{2}$}}
\put(4.7,1.5){{\small $\tilde{\chi}^0_{1}$}}
\put(2.1,3.9){{\small $\tilde{\ell}_{L,R}$}}
\end{picture}}
\end{center}
\end{minipage}
\end{center}
\vspace{-1.7cm}
\caption{\label{Fig:FeynDecayA}Feynman diagrams for the three-body leptonic neutralino decay
$\tilde{\chi}^0_2\to \tilde{\chi}^0_1\ell^{+}\ell^{-}$.}
\end{figure}

We consider spectra where the second step in the cascade decay chain is the three-body decay of the neutralino,
$\tilde{\chi}^0_2 \to \tilde{\chi}^0_1 \ell^+ \ell^-$ (cf.\ Appendix~\ref{sect:neutdecay}).  The neutralino decay
occurs via $Z^0$ exchange in the direct channel and via slepton $\tilde{\ell}_{L,R}$ exchanges in the
$t$ and $u$ channels, as can be seen in Fig.~\ref{Fig:FeynDecayA}.  It is sensitive to CP-violating
supersymmetric couplings and its structure has been studied in detail 
in \cite{Choi:2005gt,Bartl:2004jj,MoortgatPick:1999di}, cf.\ Appendix~\ref{sec:lagrangian-couplings}. The phase $\phi_{1}$ affects the masses of $\tilde \chi_1^0$ and $\tilde{\chi}_2^0$, 
as well as its couplings and decay rates. 

Using the narrow width approximation and the formalism of~\cite{Haber:1994pe},
the squared amplitude $|T|^2$ of the full process can be factorised into the 
processes of production $pp\to \tilde{g} \tilde{q}_L$ and the subsequent decays 
$\tilde{q}_L \to q \tilde{\chi}^0_2$, 
$\tilde{\chi}^0_2\to\tilde{\chi}^0_1\ell^+\ell^-$ taking into account the full spin information of the decaying $\tilde{\chi}^0_2$. For the purpose of analysing neutralino decays we do not need to include decays of the gluino produced with a squark but they will be needed later for momentum reconstruction, see Sec.~\ref{sec:MomRec}. We apply the narrow-width approximation for the
masses of the intermediate particles, $\tilde{q}_L$ and $\tilde{\chi}^0_2$, which is appropriate since
the widths of the respective particles are in all cases much smaller than their masses and the mass differences between them are large enough, see Appendix~\ref{sec:kinematics}. The squared amplitude can then be expressed in the form
\begin{eqnarray}
|T|^2 &=& 4 |\Delta(\tilde{q}_L)|^2 |\Delta(\tilde{\chi}^0_2)|^2 P(\tilde{g}\tilde{q}_L) \Big{\{} P(\tilde{\chi}^0_2)
 D(\tilde{\chi}^0_2) +
            \sum^3_{a=1}\Sigma^a_P(\tilde{\chi}^0_2)
                    \Sigma^a_D(\tilde{\chi}^0_2)\Big\},
\label{Tsquared}
\end{eqnarray}
where $a=1,2,3$ refers to the polarisation states of the neutralino $\tilde{\chi}^0_2$, which are described by the polarisation vectors $s^a(\tilde{\chi}^0_2)$, given in Appendix~\ref{sect:squarkdecay}.  In addition,
\begin{itemize}
\item $\Delta(\tilde{q}_L)$ and $\Delta(\tilde{\chi}^0_2)$ are the
  propagators of the intermediate particles which lead to the factors
  $E_{\tilde{q}_L}/m_{\tilde{q}_L}\Gamma_{\tilde{q}_L}$ and
  $E_{\tilde{\chi}^0_2}/m_{\tilde{\chi}^0_2}\Gamma_{\tilde{\chi}^0_2}$ in the narrow-width approximation,
\item $P(\tilde{g}\tilde{q}_L)$, $P(\tilde{\chi}^0_2)$ and $D(\tilde{\chi}^0_{2})$ are the terms in the
  production and decay that are independent of the polarisations of the decaying 
  neutralino,  whereas
\item $\Sigma^a_P(\tilde{\chi}^0_{2})$ and $\Sigma^a_D(\tilde{\chi}^0_{2})$
  are the terms containing the correlations between production and decay spin of the $\tilde{\chi}^0_2$.
\end{itemize}

According to our choice of the polarisation vectors $s^a(\tilde{\chi}^0_2)$, see
Eq.~(\ref{eq-s1chi}) in Appendix~\ref{sect:squarkdecay}, $\Sigma^3_P/P(\tilde{\chi}^0_2)$ is
the longitudinal polarisation, $\Sigma^1_P/P(\tilde{\chi}^0_2)$ is the transverse polarisation in the
production plane, and $\Sigma^2_P/P(\tilde{\chi}^0_2)$ is the polarisation perpendicular to the reference
plane of the neutralino $\tilde{\chi}^0_2$.

\subsection{Structure of the T-odd asymmetry}
\label{sec:Todd_Struc}

The tools used in this paper to study CP-violating effects are T-odd observables that are based on triple products of the momenta of the involved particles. Here we study the visible decay products of the $\tilde{q}_L$ in the form of the following triple product that can be evaluated in different reference frames:
\begin{equation}
  \label{eq:tpt}
 \mathcal{T}= \vec{p}_q \cdot (\vec{p}_{\ell^+} \times \vec{p}_{\ell^-}).
\end{equation}
The T-odd asymmetry is then defined as,
\begin{eqnarray}
\label{Asy}
\mathcal{A}_{\mathcal{T}} = 
\frac{N_{\mathcal{T}_+}-N_{\mathcal{T}_-}}{N_{\mathcal{T}_+}+N_{\mathcal{T}_-}} &=&
\frac{\int\mathrm{sign}\{ \mathcal{T}\}
  |T|^2d\,\mbox{lips}}{{\int}|T|^2d\,\mbox{lips}}\nonumber \\
 &  =& \frac{\int \mathrm{sign} \{ \mathcal{T}\} 
 \Big( \sum^3_{a=1}\Sigma^a_P(\tilde{\chi}^0_2)
                    \Sigma^a_D(\tilde{\chi}^0_2)\Big)d\,\mbox{lips}}{\int
 \Big(P(\tilde{\chi}^0_2)
 D(\tilde{\chi}^0_2)\Big)d\,\mbox{lips}}\:,
 \end{eqnarray}
where $N_{\mathcal{T}_+}$ ($N_{\mathcal{T}_-}$) are the numbers of events for which
$\mathcal{T}$ is positive (negative) and $d\,\mbox{lips}$ denotes Lorentz invariant phase space. The denominator in Eq.~(\ref{Asy}), ${\int}|T|^2d\,\mbox{lips}$, is equal to the total
cross section, namely $\sigma(p p \to
\tilde{q}_L \tilde{g} \to q \tilde{\chi}^0_1 \ell^+ \ell^- \tilde{g})$.  In the corresponding numerator
of Eq.~(\ref{Asy}), the triple-product correlations only enter via the spin-dependent terms. If the spin of the particles is neglected in the calculation the asymmetry will vanish.

The asymmetry $\mathcal{A}_{\mathcal{T}}$ can be visualised as the difference between the number of events where the observed $\vec{p}_q$ (from the decaying $\tilde{q}_L$) lies above ($N_{\mathcal{T}_+}$) or below ($N_{\mathcal{T}_-}$) the plane spanned by $ ( \vec{p}_{\ell^+},\vec{p}_{\ell^-} ) $, normalised by the total number of these events. If no complex phases are present, the $\vec{p}_q$ will on average line up with the plane and no asymmetry will be seen. The asymmetry is not a Lorentz invariant quantity but it is maximal in the rest frame of the $\tilde{\chi}^0_2$. At the LHC the momentum of the $\tilde{\chi}^0_2$ will in general be undetermined. The asymmetry can still be formed in the laboratory frame but it will be significantly diluted compared to the $\tilde{\chi}^0_2$ rest frame due to the initial particle boost that is parametrised by the parton density functions (PDFs). The dilution is due to the fact that the relative orientation of the plane spanned by $\vec{p}_{\ell^+}, \vec{p}_{\ell^-}$ and the quark momentum can change when boosting from the neutralino rest frame to the laboratory frame. Hence, the quark that was initially above the plane can instead be observed to come from below the plane in the laboratory frame. A possible solution to bypass the effects of the dilution would be to reconstruct the momentum of the $\tilde{\chi}^0_1$ and use it to find the momentum of the $\tilde{\chi}^0_2$. We will discuss this method in Sec.~\ref{sec:MomRec}.

In order to identify the T-odd
contributions, we have to identify those terms in $|T|^2$, Eq.~(\ref{Tsquared}), which
contain a triple product of the form shown in Eq.~(\ref{eq:tpt}).
Triple products follow from expressions $i\epsilon_{\mu\nu\rho\sigma}a^\mu b^\nu c^\rho
d^\sigma$, where $a$, $b$, $c$, $d$ are 4-momenta or spins of the particles involved,
which are non-zero only when the momenta are linearly independent.  The expressions
$i\epsilon_{\mu\nu\rho\sigma}a^\mu b^\nu c^\rho d^\sigma$ are imaginary and when
multiplied by the imaginary parts of the respective couplings they yield terms that
contribute to the numerator of $\mathcal{A}_{\mathcal{T}}$, Eq.~(\ref{Asy}). In our process, T-odd terms with $\epsilon$-tensors are only contained in the spin-dependent contributions in the neutralino decay, $\Sigma^a_D(\tilde{\chi}^0_j)$. 

Examining the covariant product, $\epsilon_{\mu\nu\rho\sigma}a^\mu b^\nu c^\rho
d^\sigma$, we can understand why the maximal asymmetry is seen in the rest frame of the $\tilde{\chi}^0_2$. If we expand the product in terms of the explicit momenta, Eq.~(\ref{eq:EpsExpan}), we see that in general we have triple products formed from all different combinations of the momenta of the four different particles in the covariant product:
\begin{eqnarray}
  \epsilon_{\mu\nu\rho\sigma}p_a^\mu p_b^\nu p_c^\rho p_d^\sigma = & E_a\;\overrightarrow{p_b}\cdot(\overrightarrow{p_c}\times\overrightarrow{p_d})+E_c\;\overrightarrow{p_d}\cdot(\overrightarrow{p_a}\times\overrightarrow{p_b}) \label{eq:EpsExpanLab} \\ & -E_b\;\overrightarrow{p_c}\cdot(\overrightarrow{p_d}\times\overrightarrow{p_a})-E_d\;\overrightarrow{p_a}\cdot(\overrightarrow{p_b}\times\overrightarrow{p_c}). \nonumber 
\end{eqnarray}
As we evaluate only one triple product we miss the other combinations and the asymmetry is not maximal. However, if we are in the rest frame of the $\tilde{\chi}^0_2$, the momentum components of the four vector vanish and we are now only left with the single triple product that we are interested in:

\begin{equation}
  \epsilon_{\mu\nu\rho\sigma}p_a^\mu p_b^\nu p_c^\rho p_d^\sigma \longrightarrow  m_a\;\overrightarrow{p_b}\cdot(\overrightarrow{p_c}\times\overrightarrow{p_d}).
 \label{eq:EpsExpanRest}
\end{equation}

Let us finally comment on the origin of the asymmetry, Eq.~(\ref{Asy}). Neutralinos in the process we study are polarised due to the nature of the electroweak coupling between the fermion, sfermion and neutralino, see also \cite{AguilarSaavedra:2004hu}. Therefore the momentum of the outgoing fermion is aligned with the spin of the neutralino $\tilde{\chi}^0_2$. The actual asymmetry arises solely in the neutralino $\tilde{\chi}^0_2$ decay as a result of the correlation between the neutralino spin and the momenta of outgoing leptons, consequently it requires us to take into account all spin correlations in the decay chain.

\section{Momentum reconstruction}
\label{sec:MomRec}

\subsection{Dilution effects}
\label{sec:DillEff}

\begin{figure}[t!]
\begin{picture}(16,7.5)
  \put(6.4,7.2){$\mathcal{T}=\overrightarrow{p}_q\cdot(\overrightarrow{p}_{\ell^+}\times\overrightarrow{p}_{\ell^-})$}
 \put(10.9,-0.2){$|\vec{p}_{\tilde{q}}|$ [GeV]}
 \put(2.9,7.2){\epsfig{file=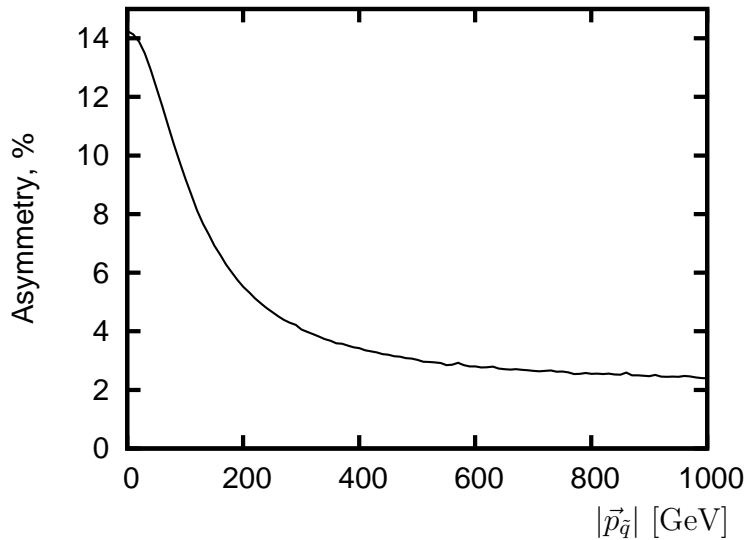,scale=0.8,angle=270}}
\end{picture}
\caption{\label{fig:MomPlot} The parton level asymmetry $\mathcal{A}_{\mathcal{T}}$, Eq.~(\protect\ref{Asy}), for the single decay chain given in Eq.~(\protect\ref{eq-decays}) as a function of the squark momentum, $|\vec{p}_{\tilde{q}}|$, in the laboratory frame. The scenario is given in Table~\protect\ref{tab:spectrum} with the phase $\phi_1=3\pi/2$. }
\vspace{0.cm}
\end{figure}

The triple product that is constructed from momenta in the laboratory frame suffers from dilution factors at the LHC due to the frame being boosted with respect to the rest frame of the $\tilde{\chi}^0_2$, for a more detailed discussion see \cite{Ellis:2008hq}. It results in a considerable reduction in the maximum asymmetry observable when we introduce the PDFs which causes an undetermined boost to the system. To illustrate the dilution, Fig.~\ref{fig:MomPlot} shows how the parton level asymmetry $\mathcal{A}_{\mathcal{T}}$, Eq.~(\protect\ref{Asy}), is diluted in the laboratory frame when we produce the $\tilde{q}$ with varying initial momenta. As the $\tilde{\chi}^0_2$ is produced in the squark decay any boost is transferred. The plot was produced with an analytical calculation for the single decay chain given in Eq.~(\protect\ref{eq-decays}). The scenario displayed in Tab.~\ref{tab:spectrum} was used but with the phase set to $\phi_1=3\pi/2$.

We see that the asymmetry is maximal, $\mathcal{A}_{\mathcal{T}}\sim14\%$, when the $\tilde{q}$ is at rest but drops to, $\mathcal{A}_{\mathcal{T}}\sim2.5\%$, when $|\vec{p}_{\tilde{q}}|\sim1$ TeV. The magnitude of the dilution clearly depends on the chosen scenario and in particular on the masses of the particles involved in the process. If it were possible to reconstruct the momentum of the $\tilde{\chi}^0_2$, we could however perform a Lorentz transformation of all the momenta and bypass the dilution factor, potentially recovering the full asymmetry.

\subsection{Reconstruction procedure}
\label{sec:RecProd}

\begin{SCfigure}[50]
  \begin{picture}(7,8.15)
 \put(1,0.3){\epsfig{file=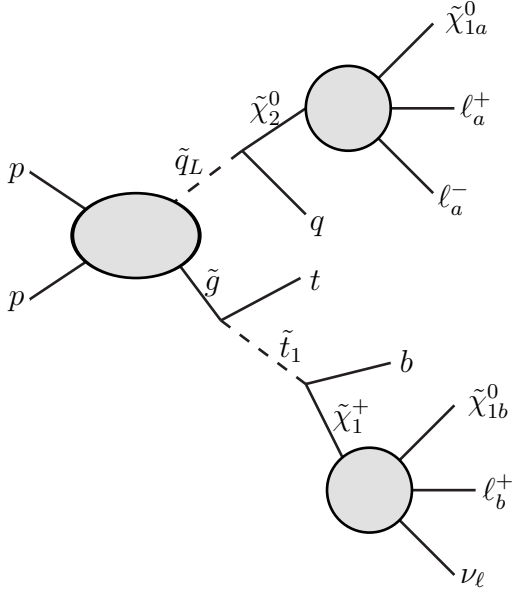,scale=0.5}}
  \put(6.9,.3){$\nu_{\ell}$}
   \put(6.6,5.3){$\ell^-_a$}
   \put(6.9,6.5){$\ell^+_a$}
    \put(6.7,7.7){$\tilde{\chi}^0_{1a}$}
    \put(0.9,5.7){$p$}
   \put(0.9,4){$p$}
   \put(3.5,4.2){$\tilde{g}$}
    \put(3.1,5.8){$\tilde{q}_L$}
     \put(4.9,5){$q$}
     \put(4.1,6.5){$\tilde{\chi}^0_2$}
     \put(4.9,4.2){$t$}
     \put(4.5,3.3){$\tilde{t}_1$}
     \put(5.2,2.4){$\tilde{\chi}^+_1$}
     \put(6.1,3.1){$b$}
     \put(7,2.6){$\tilde{\chi}^0_{1b}$}
     \put(7.2,1.4){$\ell^+_b$}
 \end{picture}
   \hspace{0.8cm}   
   \caption{\label{fig:FullDiagram} The process studied for momentum reconstruction.}
\end{SCfigure}

In the mSUGRA scenario that we have chosen to study in this paper (see Section~\ref{sec:scenario}) the full reconstruction of the event is made possible by considering the decay chains of both the particles produced in the hard collision. We include all the particles coming from both the $\tilde{q}_L$ and the $\tilde{g}$ because there are not enough kinematic constraints to perform reconstruction if only the $\tilde{q}$ chain is considered. Therefore we exclusively consider the production of $\tilde{q}_{L}$ and $\tilde{g}$ and their subsequent decays, see Fig.~\ref{fig:FullDiagram}. The $\tilde{q}_L$ decay chain will be the same as considered in Eq.~(\ref{eq-decays}) and the $\tilde{g}$ will decay as follows:
\begin{equation}\label{eq:gluinodecay}
 \tilde{g} \to \tilde{t} + t \to \tilde{\chi}^+_1 + b + t \to \tilde{\chi}^0_1 \ell^+ \nu_{\ell} + b + t.
 \label{gludec}
\end{equation}

\noindent In many scenarios the production of $\tilde{q}_L$ along with $\tilde{g}$ is the dominant source of the first and second generation squarks at the LHC and in the considered scenarios the branching ratios for the decay chain are favourable (cf.\ Sec~\ref{sec:results}).

Assuming that all the masses in the decay chains are known, Sec.~\ref{sec:scenario}, the kinematics can be fully reconstructed using the set of invariant conditions and the measured missing transverse momentum. For our procedure we follow the methods for solving the kinematic equations very closely to those presented in \cite{Cheng:2007xv,Kawagoe:2004rz}. The novelty here is the inclusion of three-body decays of sparticles and allowing for additional missing momentum due to neutrinos in the final state. The difference lies also in using the mass constraints. For our purpose we assume that masses of sparticles are known and aim at the reconstruction of the momenta, whereas the previous studies used the above conditions to reconstruct masses.

Rather than fully reconstructing the kinematics of both decay chains, an alternative idea may be to estimate the momentum of the $\tilde{\chi}^0_2$ and then boost into this approximate frame. A formula that estimates the momentum was presented in \cite{:1999fr} and is shown below,
\begin{equation}
 \vec{p}_{\tilde{\chi}^0_2}^{\,\,\mathrm{approx}} \equiv \left(1+\frac{m_{\tilde{\chi}^0_1}}{M_{\ell\ell}}\right)\vec{p}_{\ell\ell}.
\end{equation}
This approach does not work in our study however as the approximation only becomes valid when $\vec{p}_{\tilde{\chi}^0_1}\to0$ in the rest frame of the $\tilde{\chi}^0_2$. At this kinematical endpoint the two leptons are back to back which causes the plane spanned by $\vec{p}_{l^+}$ and $\vec{p}_{l^-}$ to become badly defined. The triple product is therefore small and inaccurate leading to an asymmetry that is close to zero. Therefore the approach is not valid in our case.

\vspace{0.5cm}

In our process the following invariant equations can be formed.
\begin{itemize}
 \item $\tilde{q}$ decay chain:
       \begin{eqnarray}
	 m^2_{\tilde{\chi}^0_1} & = & (P_{\tilde{\chi}^0_{1a}})^2 \label{eq:chimass},\\
         m^2_{\tilde{\chi}^0_2} & = & (P_{\tilde{\chi}^0_{1a}}+P_{\ell^+_a}+P_{\ell^-_a})^2,\\ 
	m^2_{\tilde{q}} & = & (P_{\tilde{\chi}^0_2}+P_{q})^2.
	\end{eqnarray}
\item $\tilde{g}$ decay chain:
	\begin{eqnarray}
	  m^2_{\tilde{\chi}^+_1} & = & (P_{\tilde{\chi}^0_{1b}}+P_{\nu_{\ell}}+P_{\ell^+_b})^2 \label{eq:gME}\,,\\
	  m^2_{\tilde{t}} & = & (P_{\tilde{\chi}^+_1}+P_b)^2, \\
	 m^2_{\tilde{g}} & = & (P_{\tilde{t}}+P_{t})^2, 
	\end{eqnarray}
where $P$ denote 4-momenta of respective particles, and where necessary we label particles coming from squark and gluino decays with subscripts $a$ and $b$, respectively.

\item We also have the missing transverse momentum constraint: 
	\begin{eqnarray}
	 \overrightarrow{p}^T_{miss} & = & \overrightarrow{p}^T_{\tilde{\chi}^0_{1a}} + \overrightarrow{p}^T_{\tilde{\chi}^0_{1b}} + \overrightarrow{p}^T_{\nu_{\ell}}\,.  \label{eq:MET}
          \end{eqnarray}
\item We combine the momenta of $\tilde{\chi}^0_{1b}$ and $\nu_{\ell}$ coming from the gluino together as it is impossible to resolve these two particles:
	\begin{equation}
	 P_{\tilde{g}ME} = P_{\tilde{\chi}^0_{1b}}+P_{\nu_{\ell}}\,.
	\end{equation}
\item An additional condition on the solutions is that the invariant $P^2_{\tilde{g}ME}$ has to be greater than the mass of the LSP, $m^2_{\tilde{\chi}^0_1}$:
 	\begin{equation}
 	 P^2_{\tilde{g}ME}>m^2_{\tilde{\chi}^0_1}\,. \label{eq:pgmeinv}
 	\end{equation}
We apply this condition to each solution and discard as unphysical any that does not meet it.
\end{itemize}
After expressing the momenta of intermediate particles in terms of the final state particles
\begin{eqnarray}
P_{\tilde{\chi}^0_2} & = & P_{\tilde{\chi}^0_{1a}} + P_{l^+_a} + P_{l^-_a}\,,\\
P_{\tilde{\chi}^+_1} & = & P_{\tilde{g}ME} + P_{\ell^+_b} \,,\\
P_{\tilde{t}} & = & P_b + P_{\tilde{\chi}^0_{1b}}+P_{\ell^+_b}+P_{\nu_{\ell}} \,,
\end{eqnarray}
we now have a total of eight equations, Eq.~(\ref{eq:chimass}) - (\ref{eq:MET}), and eight unknowns: 
\begin{eqnarray}
& \bigg(E_{\tilde{\chi}^0_{1a}},p_{\tilde{\chi}^0_{1a}}(x),p_{\tilde{\chi}^0_{1a}}(y),p_{\tilde{\chi}^0_{1a}}(z)\bigg)\,,\\
& \bigg(E_{\tilde{g}ME},p_{\tilde{g}ME}(x),p_{\tilde{g}ME}(y),p_{\tilde{g}ME}(z)\bigg)\:.
\end{eqnarray}

 In principle the system can be solved to find $P_{\tilde{\chi}^0_1}$ and $P_{\tilde{g}ME}$. Equations Eq.~(\ref{eq:chimass}) and Eq.~(\ref{eq:gME}) are quadratic in $P_{\tilde{\chi}^0_1}$ and $P_{\tilde{g}ME}$ respectively, so we consider these last. Using on-shell conditions, quadratic terms in the remaining equations can be removed giving 6 linear equations, therefore we can simply use a matrix to give us solutions in terms of the energies $E_{\tilde{\chi}^0_1}$ and $E_{\tilde{g}ME}$. We first expand $\overrightarrow{p}_{\tilde{\chi}^0_1}$ and $\overrightarrow{p}_{\tilde{g}ME}$ in terms of other momenta contained in the respective decay chains:
	\begin{eqnarray}
	 \overrightarrow{p}_{\tilde{\chi}^0_{1a}} & = & A\overrightarrow{p}_{\ell^+_a} + B\overrightarrow{p}_{\ell^-_a} + C\overrightarrow{p}_{q}\:, \label{eq:ABC} \\  
	\overrightarrow{p}_{\tilde{g}ME} & = & D\overrightarrow{p}_{\ell^+_b} + E\overrightarrow{p}_{b} + F\overrightarrow{p}_{t}\:.\label{eq:DEF}
          \end{eqnarray}
We can now form the system of 6 linear equations for unknowns $A$-$F$ with $E_{\tilde{\chi}^0_{1a}}$ and $E_{\tilde{g}ME}$ as free parameters
\begin{equation}
  \label{eq:MomMatEq}
  \mathcal{M}\left(
    \begin{array}{c}
      A \\ B \\ C \\ D \\ E \\ F
    \end{array} \right)= \left(
    \begin{array}{c}
       \frac{1}{2}(m^2_{\tilde{\chi}^0_1}-m^2_{\tilde{\chi}^0_2}) + P_{\ell^+_a}\cdot P_{\ell^-_a} + E_{\tilde{\chi}^0_{1a}}(E_{\ell^+_a}+E_{\ell^-_a})    \\
       \frac{1}{2}(m^2_{\tilde{\chi}^0_2}-m^2_{\tilde{q}}) + P_{\ell^+_a}\cdot P_{q} + P_{\ell^-_a}\cdot P_{q}  + E_{\tilde{\chi}^0_{1a}}E_{q}\\
       \frac{1}{2}(m^2_{b}+m^2_{\tilde{\chi}^+_1}-m^2_{\tilde{t}}) + P_{\ell^+_b}\cdot P_{b} +E_{\tilde{g}ME}E_b \\
       \frac{1}{2}(m^2_{\tilde{t}}+m^2_{t}-m^2_{\tilde{g}}) + P_{\ell^+_b}\cdot P_{t} + P_{b}\cdot P_{t} +E_{\tilde{g}ME}E_t \\
       p^T_{miss}(x) \\
       p^T_{miss}(y) \\
    \end{array} \right)\:,
\end{equation}
where
\begin{eqnarray}
 &&\hspace*{-1.5cm}  \mathcal{M} = \left(
    \begin{array}{cccccc}
       (\orp_{\ell^+_a}+\orp_{\ell^-_a}) \orp_{\ell^+_a} & (\orp_{\ell^+_a}+\orp_{\ell^-_a}) \orp_{\ell^-_a} & (\orp_{\ell^+_a}+\orp_{\ell^-_a}) \orp_{q} & 0 & 0 & 0  \\
       \orp_{\ell^+_a} \orp_{q} & \orp_{\ell^-_a} \orp_{q} & \orp_{q} \orp_{q} & 0 & 0 & 0 \\
       0 & 0 & 0 & \orp_b \orp_{\ell^+_b} & \orp_b \orp_{b} & \orp_b \orp_t \\
       0 & 0 & 0 & \orp_t \orp_{\ell^+_b} & \orp_b \orp_{t} & \orp_t \orp_t \\
       p_{\ell^+_a}(x) & p_{\ell^-_a}(x) & p_q(x) & p_{\ell^+_b}(x) & p_b(x) & p_t(x)  \\
       p_{\ell^+_a}(y) & p_{\ell^-_a}(y) & p_q(y) & p_{\ell^+_b}(y) & p_b(y) & p_t(y)  \\
    \end{array} \right)\nonumber \\ 
 &&
\end{eqnarray}
The matrix $\mathcal{M}$ can then be diagonalised to give solutions for each of the momentum components of $\overrightarrow{p}_{\tilde{\chi}^0_1}$ and $\overrightarrow{p}_{\tilde{g}ME}$ in terms of $E_{\tilde{\chi}^0_{1a}}$ and $E_{\tilde{g}ME}$. These solutions are substituted into the two quadratic equations, Eq.~(\ref{eq:chimass}) and Eq.~(\ref{eq:gME}), to produce two equations of the form:
	\begin{eqnarray}
	 a_{11}E_{\tilde{\chi}^0_{1a}}^2+a_{12}E_{\tilde{\chi}^0_{1a}}E_{\tilde{g}ME} +a_{22}E_{\tilde{g}ME}^2 +a_1E_{\tilde{\chi}^0_{1a}} +a_2E_{\tilde{g}ME} +a \equiv F_A & = &  0\,, \label{eq:aEs}\\
	 b_{11}E_{\tilde{\chi}^0_{1a}}^2+b_{12}E_{\tilde{\chi}^0_{1a}}E_{\tilde{g}ME} +b_{22}E_{\tilde{g}ME}^2 +b_1E_{\tilde{\chi}^0_{1a}} +b_2E_{\tilde{g}ME} +b \equiv F_B & = & 0\,, \label{eq:bEs}
	\end{eqnarray}
where the coefficients $a_{ij}$, $a_i$, $a$ and $b_{ij}$, $b_i$, $b$ are functions only of masses and measured momenta.
We use,
	\begin{equation}
 	  F_A-\frac{a_{11}}{b_{11}}\times F_B =0
	\end{equation}
to produce the linear equation for $E_{\tilde{\chi}^0_{1a}}$,
	\begin{equation}
 	  E_{\tilde{\chi}^0_{1a}}=\frac{a_{11}b-a\,b_{11}-a_2b_{11}E_{\tilde{g}ME}+a_{11}b_2E_{\tilde{g}ME}- a_{22}b_{11}E^2_{\tilde{g}ME}+a_{11}b_{22}E^2_{\tilde{g}ME}}{-a_{11}b_1+a_1b_{11}+a_{12}b_{11}E_{\tilde{g}ME} -a_{11}b_{12}E_{\tilde{g}ME}}\,. \label{eq:EChi}
	\end{equation}
This result can then be substituted into Eq.~(\ref{eq:aEs}) to obtain a quartic equation of the following form:
	\begin{equation}
	 Q_4E_{\tilde{g}ME}^4+Q_3E_{\tilde{g}ME}^3+Q_2E_{\tilde{g}ME}^2+Q_1E_{\tilde{g}ME}+Q_0=0\,,\label{eq:quartic}
	\end{equation}
where the various $Q$'s are functions of the $a$'s and $b$'s in Eqs.~(\ref{eq:aEs}) and (\ref{eq:bEs}).

\subsection{Discussion of graphical solutions}

Analysing the roots of the quartic equations, Eq.~(\ref{eq:quartic}), we select the solutions that are real and discard the solutions that contain imaginary parts. The selected roots are substituted into Eq.~(\ref{eq:EChi}) to find the corresponding solutions for $E_{\tilde{\chi}^0_{1a}}$. Using the values of $E_{\tilde{g}ME}$ and $E_{\tilde{\chi}^0_{1a}}$ together with the diagonalised matrix we can now calculate $A, B, C, D, E, F$, see Eqs.~(\ref{eq:ABC}),(\ref{eq:DEF}) and hence the components of $\overrightarrow{p}_{\tilde{g}ME}$ and $\overrightarrow{p}_{\tilde{\chi}^0_{1a}}$.

\begin{figure}[t]
\begin{picture}(16,8)
\put(4.1,0.4){\epsfig{width=0.5\textwidth,file=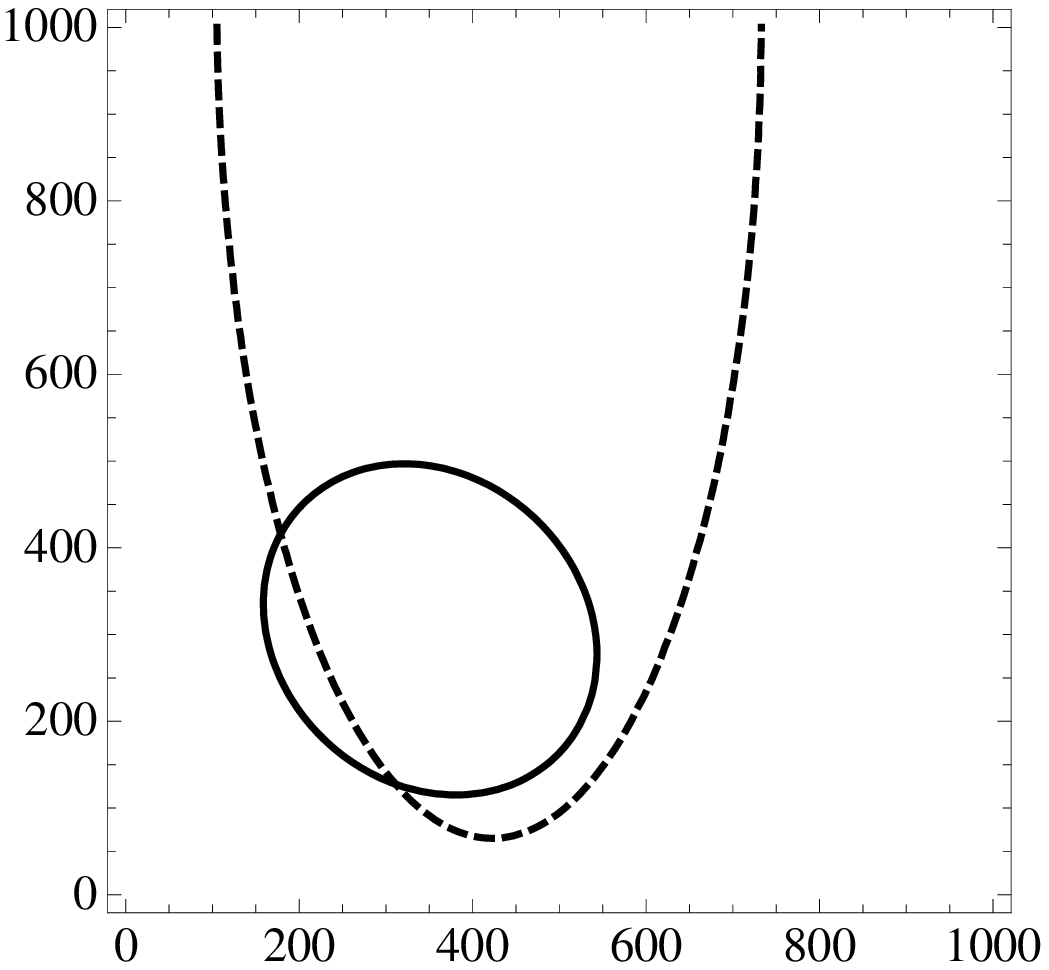}}
\put(9.8,-0.){$E_{\tilde{g}ME}$ [GeV]} \put(2.2,7){$E_{\tilde{\chi}^0_{1a}}$ [GeV]}
\end{picture}
\caption{Ellipses in the $E_{\tilde{\chi}^0_{1a}}$ and $E_{\tilde{g}ME}$ plane corresponding to Equations~(\protect\ref{eq:aEs}) (solid line) and (\protect\ref{eq:bEs}) (dashed line) for one sample event. Out of the two real solutions the lower right one gives the correct momenta of the original event. \label{fig:contours}}
\end{figure}

In general, taking into account multiple roots, Eq.~(\ref{eq:quartic}) has 4 solutions. Thus we would have 4 real, 2 real and 1 complex pair or 2 complex pairs of roots. Only the real roots can yield physical solutions, therefore for each event we normally expect real solutions. As Eq.~(\ref{eq:quartic}) is derived from Eqs.~(\ref{eq:aEs}) and (\ref{eq:bEs}) they share the same set of solutions. Both Equations~(\ref{eq:aEs}) and (\ref{eq:bEs}) are polynomials of degree 2 in $E_{\tilde{\chi}^0_{1a}}$ and $E_{\tilde{g}ME}$ so they correspond to degree 2 curves in the $E_{\tilde{\chi}^0_{1a}}$ and $E_{\tilde{g}ME}$ plane (ellipses, hyperbolas or parabolas). The intersection points solve simultaneously both equations and at the same time the quartic Equation~(\ref{eq:quartic}). A graphical solution to this set of equations is shown in Fig.~\ref{fig:contours} for one sample event. In this example we have two real solutions, of which only one corresponds to the actual momenta in the event.

In the realistic physical case one has to include uncertainties on measured momenta and masses. This will of course affect above equations and solutions. Smearing of momenta will typically result in shifting of the curves shown in Fig.~\ref{fig:contours}. Therefore it is possible that one can get two additional solutions or no real solutions at all. The consequences of experimental effects on our analysis will be discussed in Sec.~\ref{sec:expfactors}.

\subsection{Practical approach}
\label{sec:PracApproach}

In general more than one solution to the above system will remain depending upon the resolvent. In some cases Eq.~(\ref{eq:pgmeinv}) reduces the number of solutions further. For the application of this study we therefore calculate the triple product, Eq.~(\ref{eq:tpt}), for each individual solution. If all solutions produce the same sign triple product we keep the event and use it in our asymmetry. If any of the solutions disagrees on the sign of the triple product we will discard the whole event and it will not contribute. This method has the disadvantage that we lose events and therefore statistical significance. However, we note that it does not introduce any bias, i.e.\ it can be checked that the asymmetry for accepted events remains the same. An alternative method we propose is to weight each solution with matrix element and phase space factors to select the solutions with the highest probability of being correct. A possible advantage of this method is that no events will be lost and this idea is followed up in \cite{newpaper}.

An additional issue when trying to complete the momentum reconstruction procedure are combinatorial problems when trying to assign the measured momenta to the correct particle in the given event. If we take the leptons as an example, we know that opposite sign, same flavour leptons must come from the $\tilde{\chi}^0_2$. However, it is possible that a same flavour lepton could be produced from the $\tilde{\chi}^+_1$ in the opposite decay chain and be confused with those coming from the $\tilde{\chi}^0_2$. In order to assign the leptons correctly to the decay chains one can use the conditions for invariant masses. In the squark decay chain a useful observable is the invariant mass of two leptons and in the gluino decay chain one can use the invariant mass of the $b$-jet from the stop decay and the lepton from chargino decay. The end points of these distributions are given by
\begin{eqnarray}
M_{\ell^+_a \ell^-_a} &<& m_{\tilde{\chi}^0_2} - m_{\tilde{\chi}^0_1}\label{eq:lepinv}\,,\\
M_{b \ell_b^+} &<& \sqrt{\frac{ (m^2_{\tilde{t}_1} - m^2_{\tilde{\chi}^+_1} - m_b^2) (m^2_{\tilde{\chi}^0_2} - m^2_{\tilde{\chi}^0_1})}{m_{\tilde{\chi}^+_1} m_{\tilde{\chi}^0_2} } } \label{eq:blinv}\:.
\end{eqnarray}
It turns out that only a small fraction of the events, around 5\% in our scenario, satisfies both conditions simultaneously for the two possible assignments of the leptons in question. Moreover, if we run the momentum reconstruction algorithm, we find that physically acceptable solutions are only found in roughly 10\% of events where a wrong assignment has occurred. These conditions act as a strong discriminant, therefore we do not expect lepton combinatorics to be a relevant issue for this study.

Another possible approach to this problem would be the subtraction of the opposite-sign opposite-flavour (OSOF) lepton pairs. In that case the quantities $N_{\mathcal{T}_+}$ and $N_{\mathcal{T}_-}$ in Eq.~(\ref{Asy}) would be defined as
\begin{eqnarray}
 N_{\mathcal{T}_+} &=& [N_{e^+e^-} + N_{\mu^+\mu^-} - N_{e^+\mu^-} - N_{e^-\mu^+}]_{\mathcal{T}_+}\,, \label{eq:NTplus}\\
 N_{\mathcal{T}_-} &=& [N_{e^+e^-} + N_{\mu^+\mu^-} - N_{e^+\mu^-} - N_{e^-\mu^+}]_{\mathcal{T}_-} \,. \label{eq:NTminus}
\end{eqnarray}
In Eqs.(\ref{eq:NTplus}) and (\ref{eq:NTminus}) one adds the number of events with a positive (negative) triple product using $e^+ e^-$ pairs (and the jet) to the number of events with a positive (negative) triple product using $\mu^+ \mu^-$ pairs and subtract the number of events with positive triple product using $e^+\mu^-$ and $e^-\mu^+$ pairs. For the combinatorial background one gets equal rates for same flavour and opposite flavour lepton pairs since the leptons are uncorrelated. Hence, this ``flavour subtraction'' procedure removes (up to statistical fluctuations) all combinatorial background resulting from the lepton pairing problem and also all background coming from $\tilde{\chi}_2^0 \to \tilde{\chi}_1^0 \tau^+ \tau^-$.

\section{Numerical results}
\label{sec:results}

\subsection{Chosen scenario: spectrum and decay modes}
\label{sec:scenario}

In order to study the experimental prospects of measuring CP-violating effects at the LHC we use an MSSM scenario derived from mSUGRA parameters defined at the GUT scale, as shown in Tab.~\ref{tab:spectrum}. This scenario has been already used to analyse the properties of the neutralino sector in \cite{Choi:2005gt}. The values of the parameters at the electroweak scale have been derived using the RGE code \texttt{SPheno}~\cite{Porod:2003um}. Masses of the coloured particles are of order 500~GeV, apart from the light stop $\tilde{t}_1$, which has a mass of 171.0~GeV. The lightest supersymmetric particle (LSP) is the lightest neutralino with a mass of 78.1~GeV. The second neutralino and the light chargino have masses around 150~GeV, whereas the sleptons are around 200~GeV. Therefore we note that the two-body decay channels of $\tilde{\chi}^0_2$ and $\tilde{\chi}^\pm_1$ are closed and this gives a good opportunity to study CP-violation effects in their three-body decay modes. Details of the mass spectrum can be found in Tab.~\ref{tab:spectrum}. The values of the gaugino mass parameters reproduce the given spectrum in the case when all the CP phases are set to 0. In order to generate CP-violating effects in the following we will attribute a non-zero phase $\phi_1$ to the bino mass parameter,
\begin{eqnarray}
M_1 = |M_1| e^{i \phi_1}\, , \qquad 0 \leq \phi_1 < 2\pi \,,
\end{eqnarray}
while keeping the absolute value $|M_1|$ fixed as given in Tab.~\ref{tab:spectrum}.

\begin{table}[t!]\renewcommand{\arraystretch}{1.3}
\begin{center}
\begin{tabular}{|c|c||c|c||c|c|} \hline
Parameter & Value & Particle & Mass [GeV] & Particle & Mass [GeV] \\ \hline\hline
$m_0$ & 150 GeV & $\tilde{g}$ & 496.5 & $\tilde{\chi}_1^0$ & 78.1 \\
$m_{1/2}$ & 200 GeV & $\tilde{d}_L$ & 484.1 & $\tilde{\chi}_2^0$ & 148.4 \\
$A_0$ & -650 GeV & $\tilde{d}_R$ & 466.4 & $\tilde{\chi}_1^\pm$ & 148.2 \\
$\tan\beta$ & 10 & $\tilde{u}_L$ & 477.9 & $\tilde{\chi}_2^\pm$ & 436.0 \\ \cline{5-6}
$\mathrm{sign}\: \mu$ & $+$ & $\tilde{u}_R$ & 465.9 & $\tilde{e}_L$ & 207.5 \\ \cline{1-2}
$|M_1|$ & 80.5 GeV & $\tilde{b}_1$ & 397.2 & $\tilde{e}_R$ & 173.1 \\
$M_2$ & 153.3 GeV& $\tilde{b}_2$ & 462.6 & $\tilde{\nu}_e$ & 192.0 \\ \cline{5-6}
$M_3$ & 484.6 GeV& $\tilde{t}_1$ & 171.0 & $\tilde{\tau}_1$ & 149.4 \\
$\mu$ & 419.0 GeV& $\tilde{t}_2$ & 498.0 & $\tilde{\tau}_2$ & 212.5 \\
 &  &  &  & $\tilde{\nu}_\tau$ & 187.2 \\ \hline
\end{tabular}
\caption{mSUGRA input parameters at the GUT scale, MSSM parameters and particle masses in GeV from \texttt{SPheno 2.2.3}~\cite{Porod:2003um} for $\phi_1 = 0$ and with $m_t = 171$~GeV. \label{tab:spectrum}}
\end{center}
\end{table}

Although we chose a specific scenario our method is applicable in a wider range of parameter points. For the decay chain that we concentrate on, we require the three-body decay of the $\tilde{\chi}^0_2$ which places the following constraints on the SUSY masses,
\begin{eqnarray}
  m_{\tilde{\chi}^0_2}<m_{\tilde{\ell}}\label{eq:lepcons}\,,\\  
  m_{\tilde{\chi}^0_2}-m_{\tilde{\chi}^0_1}<m_Z.\label{eq:Zcons}
\end{eqnarray}
Equation~(\ref{eq:lepcons}) ensures that the decay, $\tilde{\chi}^0_2\to \tilde{\ell}\ell$ cannot occur whereas Eq.~(\ref{eq:Zcons}) forbids the decay $\tilde{\chi}^0_2\to Z\tilde{\chi}^0_1$.

To complete momentum reconstruction from the gluino side, other decay modes can contribute with regard to the scenario presented here but their impact on the final result would be scenario dependent. For example if the decay $\tilde{t}\to t\tilde{\chi}^0_2$ became kinematically open the momentum reconstruction would actually become easier as there would be no neutrino in the final state and the system would become over-constrained. However, there would now be one additional top and one additional lepton in the final state that may cause new combinatorical difficulties.

As already mentioned, the highest production rate is typically obtained for coloured final states containing squarks and gluinos at the LHC. In our scenario, where their masses are not very heavy, the total cross section for production of strongly interacting supersymmetric particles reaches 140~pb, see Tab.~\ref{tab:CrossSec}. For our purpose we will be interested in the inclusive production of left squarks and the associated production of left squark and gluino\footnote{Triple product asymmetry in right squarks decay chain has the opposite sign as compared to the left squarks (and the same sign as that of left anti-squarks), however in our case where the coupling of the right squark to the second lightest neutralino is suppressed due to small bino component of $\tilde{\chi}^0_2$, this contribution remains negligible.}. As was mentioned in Sec.~\ref{sec:introduction}, anti-squarks give a CP asymmetry with exactly the opposite sign to squarks. We note however that the inclusive cross section for left squark production is almost a factor of 4 larger than the cross section for left anti-squark production. This is a direct consequence of the fact that we have two protons in the initial state for which the abundance of quarks is significantly higher than of anti-quarks. Since we cannot distinguish experimentally squarks and anti-squarks of the first two generations this fraction of anti-squarks will cause some dilution in the asymmetry that will be however compensated for by a very high production rate. A similar situation occurs for associated squark gluino production for which the ratio of squarks to anti-squarks is 18.2~pb to 3.1~pb, see Tab.~\ref{tab:CrossSec}. Together with left squarks we also have the production of right squarks at a comparable rate. However, since the latter decay almost exclusively to the lightest neutralino in our scenario, as shown in Tab.~\ref{tab:brs}, they do not give rise to the CP-odd asymmetry.

\begin{table}[t!]
\renewcommand{\arraystretch}{1.3}
\begin{center}
\begin{tabular}{|c|c|c|} \hline\hline
 Produced Particles & \multicolumn{2}{c|}{Cross Section (pb)} \\ \hline \hline
 At least one coloured SUSY particle. & \multicolumn{2}{c|}{148} \\ \hline
 At least one $\tilde{g}$. & \multicolumn{2}{c|}{58.8} \\  \hline
 $\tilde{q}_L\tilde{q}_L+\tilde{q}_L\tilde{q}_L^*+\tilde{q}_L\tilde{q}_R+\tilde{q}_L\tilde{q}_R^*+\tilde{q}_L\tilde{g}$ & \multicolumn{2}{c|}{30.0} \\ \hline
 $\tilde{q}^*_L\tilde{q}_L+\tilde{q}^*_L\tilde{q}_L^*+\tilde{q}^*_L\tilde{q}_R+\tilde{q}^*_L\tilde{q}_R^*+\tilde{q}^*_L\tilde{g}$ & \multicolumn{2}{c|}{8.3} \\ \hline
 $\tilde{q}_L \tilde{g}$ & \multicolumn{2}{c|}{18.2} \\ \hline
 $\tilde{q}^*_L \tilde{g}$ & \multicolumn{2}{c|}{3.1} \\ \hline 
 $\tilde{t}_1 \tilde{t}^*_1$ & \multicolumn{2}{c|}{66.3} \\ \hline 
\end{tabular}\\[0.5ex]
\caption{Leading order cross sections at $\sqrt{s} = 14\mathrm{TeV}$ for direct production of various particles from \texttt{Herwig++ 2.3.2}~\cite{Bahr:2008tx,Bahr:2008pv} using \texttt{MRST 2004LO} PDF set~\cite{Martin:2007bv}. $\tilde{q}$ stands for squarks of the first and second generation. \label{tab:CrossSec}}
\end{center}
\end{table}

Following the production process we include subsequent decays of squarks and gluinos. The dominant decay mode of the gluino is to the light stop and the top with a branching ratio of $BR=53.8\%$. The light stop then decays almost exclusively to $\tilde{\chi}^\pm_1 b$. Left squarks decay mainly to the light chargino and the second lightest neutralino with branching fractions of $65\%$ and $33\%$ respectively. Finally we consider the decays of chargino $\tilde{\chi}^\pm_1$ and neutralino $\tilde{\chi}^0_2$. Leptonic decays constitute in total $BR=61\%$ for the chargino and $BR=68\%$ for the neutralino decay modes. For decays into light leptons we have $BR=24.3\%$ and $BR=9\%$ for chargino and neutralino, respectively. The most interesting are the neutralino decays to electrons and muons that are used to construct the CP-sensitive triple product, Eq.~(\ref{eq:tpt}). The summary of the relevant branching ratios can be found in Tab.~\ref{tab:brs}. It may be noted that in the scenario presented, the branching ratio $\tilde{\chi}^0_2\to\tilde{\chi}^0_1\tau^+\tau^-$ is unusually high (59.3\%). This is due to the masses of the $\tilde{\chi}^0_2$ and $\tilde{\tau}$ being very close and hence the kinematic factors are favourable. The particular branching ratio is simply a coincidence in the scenario chosen however and is not a required feature for our study.

\begin{table}\renewcommand{\arraystretch}{1.3}
\begin{center}
\begin{tabular}{|c|c||c|c||c|c|} \hline
Decay & BR & Decay & BR & Decay & BR \\ \hline
$\tilde{g} \to \tilde{t}_1 \bar{t} + \mathrm{c.c.}$ & 53.8\% & $\tilde{q}_L \to \tilde{\chi}_1^\pm q $ & 65\% & $\tilde{t}_1 \to \tilde{\chi}_1^+ b $ & 98.1\% \\ 
$\tilde{g} \to \tilde{b}_1 \bar{b} + \mathrm{c.c.}$ & 26.6\% & $\tilde{q}_L \to \tilde{\chi}_2^0 q $ & 33\%  & $\tilde{t}_1 \to \tilde{\chi}_2^0 c $ & 1.6\% \\ \cline{5-6}
$\tilde{g} \to \tilde{q}_R \bar{q} + \mathrm{c.c.}$ & 11.8\% & $\tilde{q}_L \to \tilde{\chi}_1^0 q $ & 1.5\% & $\tilde{\chi}_2^0 \to \tilde{\chi}_1^0 \tau^- \tau^+$ & 59.3\% \\ \cline{3-4} 
$\tilde{g} \to \tilde{b}_2 \bar{b} + \mathrm{c.c.}$ & 3.8\% & $\tilde{\chi}_1^+ \to \tilde{\chi}_1^0 \nu_\tau \tau^+$ & 37.2\% & $\tilde{\chi}_2^0 \to \tilde{\chi}_1^0 \nu \bar{\nu}$ & 23.6\% \\ 
$\tilde{g} \to \tilde{q}_L \bar{q} + \mathrm{c.c.}$ & 3.3\% & $\tilde{\chi}_1^+ \to \tilde{\chi}_1^0 q_u \bar{q}_d$ & 38.5\% & $\tilde{\chi}_2^0 \to \tilde{\chi}_1^0 q \bar{q}$ & 8.1\% \\ \cline{1-2}
$\tilde{q}_R \to \tilde{\chi}^0_1 q $ & 98\% & $\tilde{\chi}_1^+ \to \tilde{\chi}_1^0 \nu_\mu \mu^+$ & 12.2\% & $\tilde{\chi}_2^0 \to \tilde{\chi}_1^0 e^+ e^- $ & 4.5\% \\
$\tilde{q}_R \to \tilde{\chi}^0_2 q $ & 1\% & $\tilde{\chi}_1^+ \to \tilde{\chi}_1^0 \nu_e e^+$ & 12.1\% & $\tilde{\chi}_2^0 \to \tilde{\chi}_1^0 \mu^+ \mu^- $ & 4.5\% \\ \hline
\end{tabular}
\caption{Branching ratios for the scenario defined in Tab.~\protect\ref{tab:spectrum} from \texttt{SPheno 2.2.3}~\protect\cite{Porod:2003um} for $\phi_1=0$. \label{tab:brs}}
\end{center}
\end{table}

When we vary the phase of the $M_1$ parameter, the masses and couplings in the neutralino sector are affected. First we note that the changes in the neutralino masses are negligible and smaller than the possible experimental accuracy. It turns out however, that the phase $\phi_1$ has large impact on neutralino couplings and therefore the pattern of its decay modes. The most significant change is for the light chargino decays to the LSP and a fermion pair. With increasing phase the branching ratio for $\tilde{\chi}^+_1 \to \tilde{\chi}^0_1 \nu_\tau \tau^+$ rises and eventually reaches 70\% for $\phi_1 = \pi$. At the same time the branching ratios for decays to light leptons remain roughly at the level of 10\%. As we will explain later, a chargino decay to tau, followed by a leptonic tau decay can be used for momentum reconstruction in the same way as a chargino decay to an electron and a muon. On the other hand, the decay $\tilde{\chi}_2^0 \to \tilde{\chi}_1^0 \tau^+ \tau^-$ followed by a leptonic tau decay will be a background, since it gives incorrectly reconstructed momenta. Finally, we note that in the case of neutralino $\tilde{\chi}^0_2$ decays to light leptons, the respective branching ratios increase up to 5.5\% for $\phi_1 = \pi/2$.

Since our procedure requires the sparticle masses to be known, we wish to comment on the possibility of the mass determination in our scenario. One of the standard approaches at the LHC is to study kinematic edges and endpoints of the invariant mass distributions, see e.g. \cite{Weiglein:2004hn}. In our case the masses of $\tilde{g}$, $\tilde{q}_L $, $\tilde{t}_1$, $\tilde{\chi}^0_{1,2}$ and $\tilde{\chi}^\pm_{1}$ are required. The possible invariant mass observables would include: 
\begin{itemize}
\item quark and leptons in the decay chain of $\tilde{q}_L$ followed by the squark decay to chargino or neutralino $\tilde{\chi}^0_2$, 
\item lepton pair in the neutralino $\tilde{\chi}^0_2$ decay, cf.\ Eq.~(\ref{eq:lepinv}),
\item top and bottom quarks in the gluino decay chain Eq.~(\ref{eq:gluinodecay}),
\item bottom quark and lepton in the stop decay chain, cf.\ Eq.~(\ref{eq:blinv}),
\item quark pairs in the decay of gluino to the 1st and 2nd generation squarks~\cite{Cho:2007qv}.
\end{itemize}
Fitting the above invariant masses to experimentally measured edges and endpoints should provide enough number of constraints to obtain required information. This method can be perhaps supplemented by a mass reconstruction method, see for example \cite{Nojiri:2003tu,Nojiri:2007pq,Cheng:2008mg,Cheng:2009fw,Kawagoe:2004rz,Kersting:2009ne,Kang:2009sk}. Therefore, in the following we assume that the masses are known. The impact of the mass uncertainties will be discussed in Section~\ref{sec:expfactors}.

\subsection{CP asymmetry at the parton level}\label{sec:asy-parton}

We start by discussing the dependence of the asymmetry on $\phi_1$, Eq.~(\ref{Asy}), at the partonic level studying only the decay chain presented in Eq.~(\ref{eq-decays}). In order to see a maximal effect, we place the $\tilde{q}_L$ at rest and calculate the triple product and the asymmetry in its distribution, Sec.~\ref{sec:Todd_Struc}.

\begin{figure}[t!]
\begin{picture}(16,7.5)
  \put(6.4,7.2){$\mathcal{T}=\overrightarrow{p}_q\cdot(\overrightarrow{p}_{\ell^+}\times\overrightarrow{p}_{\ell^-})$}
 \put(10.9,0.){$\phi_{1}/\pi$}
 \put(2.9,7.2){\epsfig{file=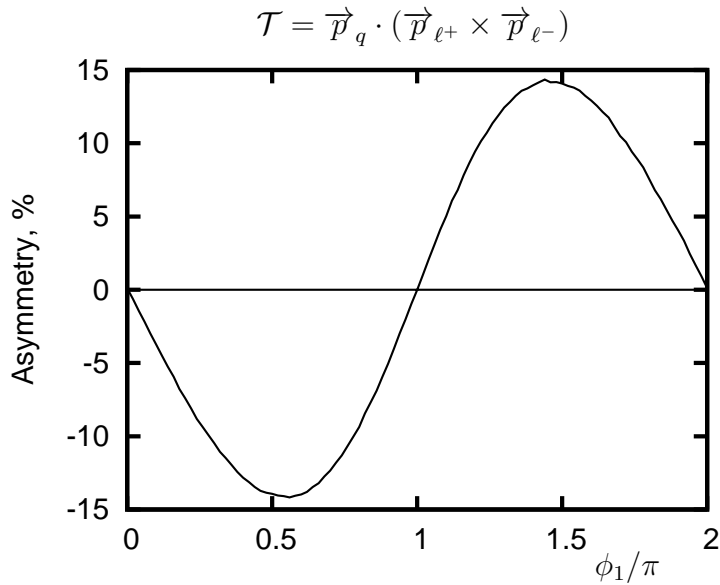,scale=0.8,angle=270}}
\end{picture}
\caption{\label{fig:PartonAsy} The parton level asymmetry $\mathcal{A}_{\mathcal{T}}$, Eq.~(\protect\ref{Asy}), for the single decay chain given in Eq.~(\protect\ref{eq-decays}) in the rest frame of $\tilde{q}_L$ as a function of $\phi_1$.   The scenario is given in Tab~\protect\ref{tab:spectrum}.}
\vspace{0.cm}
\end{figure}

We see from Fig.~\ref{fig:PartonAsy} that the asymmetry in this scenario is roughly 14\% and this occurs when the phase $\phi_{1}$ is just above $\pi/2$ and just below $3\pi/2$. The asymmetry is produced by a complex interplay between different couplings in the $\tilde{\chi}^0_2$ decay, cf. Eqs.~(\ref{eq_dszz_to})-(\ref{eq_dselel_to}), and can vary significantly between scenarios. These couplings can all have different behaviour with respect to $\phi_1$ and in other scenarios the maximum asymmetry can be seen far from $\pi/2$.

It should be noted from the previous plot that the asymmetry is obviously a CP-odd quantity that in addition to a measurement of the phase, also determines whether it lies above or below $\pi$, as seen in Fig.~\ref{fig:PartonAsy}. In comparison, using CP-even quantities, for example the mass, it is not possible to determine if the phase is above or below $\pi$. It must also be noted that these quantities have a weak dependence upon the phase and will be challenging to study. Moreover CP-even quantities alone do not give unambiguous proof of CP violation in the model, that can only be provided by CP-odd observables.

\subsection{Influence of parton distribution functions on CP asymmetries}
\label{sec:asymmetry-at-lhc}

Experimentally the situation becomes significantly more complicated since in general, particles are not produced at rest but can be heavily boosted in the laboratory frame. Our observables can be significantly reduced in size by this effect as triple product correlations induced by spin effects are largest in the rest frame of the decaying particle. Essentially, a boosted frame can make the momentum vector of the quark appear to come from the opposite side of the plane formed by $\ell^+$ and $\ell^-$. As explained in Sec.~\ref{sec:Todd_Struc} this causes a severe dilution in the asymmetry that is measured at the LHC.

There are two other dilution factors that have to be taken into account and give a further reduction in the observed asymmetry. The first one we consider is the contamination from anti-squarks $\tilde{q}_L^*$ that will be produced at the LHC along with squarks $\tilde{q}_L$. There is no way of identifying the charge of the decaying particle and consequently we need to include the effect of the $\tilde{q}_L^*$ in our analysis. Anti-squarks produce an asymmetry of the opposite sign to $\tilde{q}_L$ so if for example we had equal numbers of each of them, no overall asymmetry could be seen. However, the production cross section for $\tilde{q}_L^*$ is substantially lower than that of $\tilde{q}_L$ due to the valence quarks in the proton, see Tab.~\ref{tab:CrossSec} and at the LHC we would expect only roughly 20\% of the sample to be $\tilde{q}_L^*$.

The other background contribution is that of neutralinos decaying to a tau pair, $\tilde{\chi}_2^0 \to \tilde{\chi}_1^0 \tau^+ \tau^-$ followed by leptonic tau decays to the pair of opposite-sign same-flavour leptons. Since the branching ratio for the above decay is large, even after inclusion of leptonic tau decays there is a significant number of events faking our signal, i.e.
\begin{eqnarray}\label{eq:tautaubackg}
BR(\tilde{\chi}^0_2 \to \tilde{\chi}^0_1 \tau^+ \tau^- \to \tilde{\chi}^0_1 \ell^+ \ell^-\nu_\ell\bar{\nu}_\ell \nu_\tau \bar{\nu}_\tau) \approx 2\times0.6\times0.175\times0.175 = 3.7\%.
\label{eq:taudec}
\end{eqnarray}
compared with
\begin{eqnarray}
BR(\tilde{\chi}^0_2 \to \tilde{\chi}^0_1 \ell^+_{e,\mu} \ell^-_{e,\mu}) \approx 2\times0.045 = 9\%.
\end{eqnarray}
As the asymmetry calculated for such leptons is diluted, this background introduces a further dilution factor. However, this background can be easily removed using the flavour subtraction technique described above, Eqs.~(\ref{eq:NTplus}) and (\ref{eq:NTminus}). In addition we will see later that much of this background is removed after the application of simple selection cuts on the lepton energy and the invariant mass.

\begin{figure}[t!]
\begin{picture}(16,8)
  \put(4.6,7.4){$\mathcal{T}=\overrightarrow{p}_q\cdot(\overrightarrow{p}_{\ell^+}\times\overrightarrow{p}_{\ell^-})$, $\,\,\sqrt{s} = 14\ \mathrm{TeV}$}
 \put(10.7,0.1){$\phi_{1}/\pi$}
 \put(6.2,5.95){\tiny{20 fb$^{-1}$}}
 \put(6.2,5.25){\tiny{50 fb$^{-1}$}}
 \put(6.2,4.6){\tiny{100 fb$^{-1}$}}
 \put(9.3,6.4){3$\sigma$ Observation}
 \put(2.7,7.4){\epsfig{file=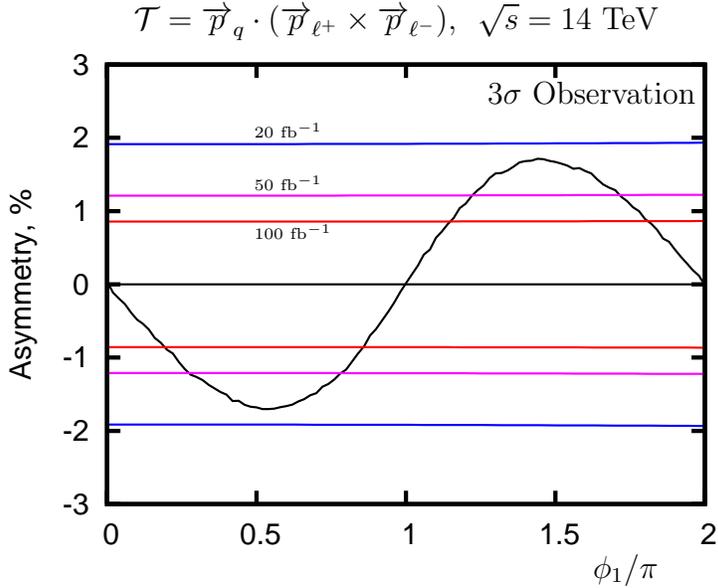,scale=0.8,angle=270}}
\end{picture}
\caption{\label{fig:PDFAsy} The parton level asymmetry $\mathcal{A}_{\mathcal{T}}$, Eq.~(\protect\ref{Asy}), in the laboratory frame with PDFs included in the analysis using the scenario shown in Tab.~\protect\ref{tab:spectrum}. The coloured lines show the size of the asymmetry needed for a $3\sigma$ observation at the given luminosity, $\mathcal{L}$=(20 fb$^{-1}$, 50 fb$^{-1}$, 100 fb$^{-1}$), assuming squarks were produced via the channels shown in rows 3 and 4 of Tab.~\protect\ref{tab:CrossSec}. All produced $\tilde{q}$ and $\tilde{q}^*$ that follow the decay chain given in Eq.~(\protect\ref{eq-decays}) are taken into account.}
\vspace{0.cm}
\end{figure}

We use the \texttt{MRST 2004LO} \cite{Martin:2007bv} PDF set in our analysis of the asymmetry and plot the integrated asymmetry $\mathcal{A_T}$ as a function of $\phi_{1}$ in Fig.~\ref{fig:PDFAsy}. We see that the inclusion of the PDFs and the $\tilde{q}_L^*$ sample reduce the asymmetry by about a factor of~8 in this scenario as compared to the result of Sec.~\ref{sec:asy-parton}. The maximum asymmetry is now $|\mathcal{A_T}|=1.7\%$. It must be noted that the dilution factor does depend on the scenario studied and changes in particular with the mass of the sparticle that is initially produced.

For the calculation of the asymmetry we included the production channels shown in rows 3 and 4 of Table~\ref{tab:CrossSec} but we only take decays of individual $\tilde{q}_L$ and $\tilde{q}_L^*$ following the decay chain in Eq.~(\ref{eq-decays}). At this point correct identification was assumed for the final state particles and no hadronisation or detector effects were included but these assumptions will be relaxed in Sec.~\ref{sec:experimental}. The only backgrounds in the study are those discussed above.

Using the total production cross sections\footnote{Note that the total rate of squark production is actually larger than 30.0~pb quoted in Table~\ref{tab:CrossSec}, as some of the subprocesses give a pair of squarks e.g.$(\tilde{q}_L\tilde{q}_L)$ and both can contribute to our analysis. The total number of squarks in the sample at the given luminosity $\mathcal{L}$ and $\sqrt{s} = 14$~TeV is therefore $33.2\times \mathcal{L}$. In case of anti-squarks the number is $8.5\times \mathcal{L}$.} for $\tilde{q}_L$ and $\tilde{q}_L^*$, and respective branching ratios from Table~\ref{tab:brs} we can now estimate
the integrated luminosity required to observe the asymmetry at the LHC with a certain significance. We assume that
$N_{\mathcal{T}_+}\,(N_{\mathcal{T}_-})$, the numbers of events where $\mathcal{T}$ is positive (negative)
as in Eq.~(\ref{Asy}), are binomially distributed, giving the following statistical 
error~\cite{Desch:2006xp}:
\begin{equation}
  \label{eq:asymmerror}
  \Delta(\mathcal{A}_T)^{\rm stat}=2\sqrt{\epsilon(1-\epsilon)/N}\,,
\end{equation}
where $\epsilon=N_{\mathcal{T}_+}/(N_{\mathcal{T}_+}+N_{\mathcal{T}_-})=\frac 12
(1+\mathcal{A}_T)$, and $N = N_{\mathcal{T}_+}+N_{\mathcal{T}_-}$ is the total number of events. This can be rearranged to
give the required number of events for a desired significance.
The horizontal lines in Fig.~\ref{fig:PDFAsy} show an estimate of the amount of luminosity required 
for a $3\sigma$ observation of a non-zero asymmetry.
In other words, an asymmetry can be seen at the 3$\sigma$ level where the asymmetry curve in Fig.~\ref{fig:PDFAsy} lies outside the luminosity band.

\subsection{Impact of momentum reconstruction on the observable CP asymmetry}
\label{sec:MomRecRes}

In order to increase the statistical significance of our CP asymmetry we investigate the possibility of reconstructing the momenta of the invisible particles in our process. In principle a perfect reconstruction would return the magnitude of the asymmetry to that where the $\tilde{q}_L$ is at rest but in reality there are additional complications with this procedure, see Sec.~\ref{sec:MomRec}. The reconstruction is performed at the partonic level with PDF's included in the production process.

\begin{figure}[t!]
\begin{picture}(16,8)
  \put(4.6,7.4){$\mathcal{T}=\overrightarrow{p}_q\cdot(\overrightarrow{p}_{\ell^+}\times\overrightarrow{p}_{\ell^-})$, $\,\,\sqrt{s} = 14\ \mathrm{TeV}$}
 \put(10.7,0.1){$\phi_{1}/\pi$}
 \put(6.2,5.2){\tiny{20 fb$^{-1}$}}
 \put(6.2,4.75){\tiny{50 fb$^{-1}$}}
 \put(6.2,4.3){\tiny{100 fb$^{-1}$}}
 \put(9.3,6.4){3$\sigma$ Observation}
 \put(2.7,7.4){\epsfig{file=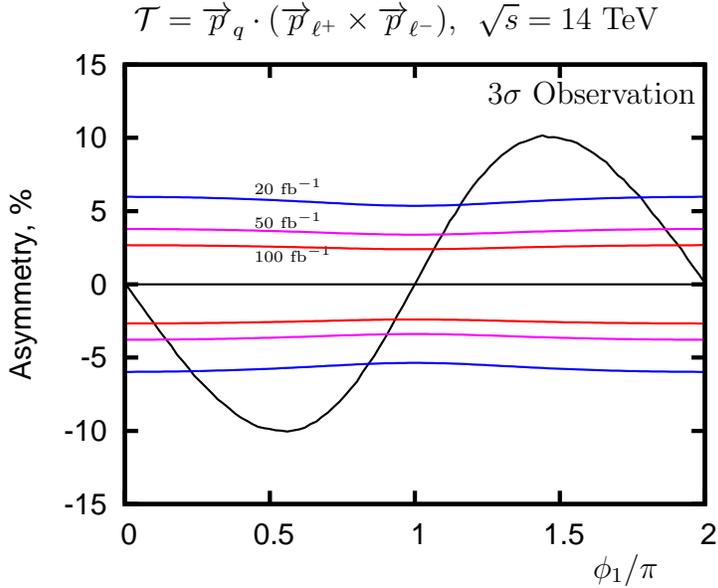,scale=0.8,angle=270}}
\end{picture}
\caption{\label{fig:MomRecAsyTau} The parton level asymmetry $\mathcal{A}_{\mathcal{T}}$, Eq.~(\protect\ref{Asy}), in the reconstructed $\tilde{q}_L$ rest frame. The coloured lines show the size of the asymmetry needed for a $3\sigma$ observation at the given luminosity, $\mathcal{L}$=(20 fb$^{-1}$, 50 fb$^{-1}$, 100 fb$^{-1}$) for the production channels $\tilde{q}_L\tilde{g}$, $\tilde{q}_L^*\tilde{g}$ Tab.~\protect\ref{tab:CrossSec}. The decay chains included are shown in Eq.~(\protect\ref{eq-decays}) and Eq~(\protect\ref{gludec}). As explained in the text correct jet and lepton assignment is assumed.}
\vspace{0.cm}
\end{figure}

Again our sample of events will contain $\tilde{q}^*_L$ which have an asymmetry of the opposite sign to that of $\tilde{q}_L$ as has already been discussed in Sec.~\ref{sec:asymmetry-at-lhc}. As we are looking exclusively at $\tilde{q}_L\tilde{g}$ and $\tilde{q}_L^*\tilde{g}$ production however when applying the momentum reconstruction, we actually have a smaller number of $\tilde{q}^*_L$ that dilute the asymmetry (15\%, see Table~\ref{tab:CrossSec}). After including this dilution factor we see our maximum asymmetry reduced from $\mathcal{A_T}\sim14\%$ to $\mathcal{A_T}\sim11\%$, see Figs.~\ref{fig:PartonAsy} and \ref{fig:MomRecAsyTau}, respectively.

To calculate the luminosity we require to see a statistically significant effect at the LHC we include the production cross sections for channels $\tilde{q}_L\tilde{g}$, $\tilde{q}_L^*\tilde{g}$ (Tab.~\protect\ref{tab:CrossSec}) and the branching ratios from both the $\tilde{q}_L$ and $\tilde{g}$ decay chains Eq.~(\ref{eq-decays}) and Eq.~(\ref{gludec}). No hadronisation or detector effects were included in this section and correct identification was assumed for the quarks. For the leptons the correct assignment is made as is explained in Sec.~\ref{sec:PracApproach}. For the jets, a detailed experimental study would be required to look at this question as we need to include hard radiation, reconstruction and $b$-tagging efficiencies. However, a cursory examination suggests that the individual jets may be resolvable. For example, the jet coming from the $\tilde{q}_L$ decay is the hardest one for a high proportion of events. We would also require three other jets to have the invariant mass $m_t$. Finally we would then require at least one $b$-jet to be tagged. We would like to state here that the only backgrounds included in the study are those of $\tilde{q}_L^*\tilde{g}$ production and the decay chain Eq.~(\ref{eq:taudec}).

 After inclusion of all branching ratios the production rate for our process drops down to $200$~fb (after hadronic top decay), which results in approximately $20\,000$ events at the integrated luminosity $\mathcal{L} = 100\ \mathrm{fb}^{-1}$. One extra point to note is that we are able to use events where the $\tilde{\chi}^\pm_1$ in the opposite decay chain produces a $\tau^\pm$. Kinematically these events are similar to our normal signal events but we have one extra neutrino $\nu_{\tau}$ that will be invisible and simply contribute to our missing transverse momenta. An additional factor that reduces the number of reconstructed and accepted events are the multiple solutions, as discussed in Sec.~\ref{sec:MomRec}. We only use events when all solutions produce the same sign triple product and this allows us to use approximately $\sim60\%$ of the events.

The horizontal lines in Fig.~\ref{fig:MomRecAsyTau} again show an estimate of the amount of luminosity required 
for a $3\sigma$ observation of a non-zero asymmetry.
It can be seen that the luminosity lines are not flat but actually give a narrower band as $\phi_1\rightarrow\pi$. This effect is caused by the branching ratios, $\tilde{q}_L\to q \tilde{\chi}^0_2$, $\tilde{\chi}^0_2\to \tilde{\chi}^0_1 \ell^+ \ell^-$ and $\tilde{\chi}^+_1\to \tilde{\chi}^0_1 \overline{\nu}_{\ell} \ell^-$ altering with the change in phase and producing more events. We see that after 20~fb$^{-1}$ of well understood LHC data it may be possible to start seeing a statistically significant effect if large phases are present. It must be noted however that for this method to be successful we will require mass measurements of the individual particles involved in the decay chains. This will obviously require significant running time and may even need the help of a precision linear collider. We will address this issue in more detail in the next section.


\section{Inclusion of experimental factors}
\label{sec:experimental}

To more realistically estimate if the study will be possible at the LHC some additional experimental factors need to be included in the analysis. For this purpose we simulate events using the Monte Carlo event generator \texttt{Herwig++ 2.3.2}, that has all the spin correlations included as required in our analysis. These events will be used in the following to perform momentum reconstruction and as a cross check for our analytic calculations.

We start with the inclusion of the selection cuts that have to be used to resolve leptons and jets and these are listed below:
\begin{eqnarray}
  E_{T}(j_1) &\geq& 100\ \mathrm{GeV}\,, \label{eq:CutJH} \\
  E_{T}(j) &\geq& 25\ \mathrm{GeV}\,,  \\
  E_{T}(\ell_{e,\mu}) &\geq& 10\ \mathrm{GeV}\,, \label{eq:leppt} \\
  M_{\ell^+\ell^-} &\geq& 20\ \mathrm{GeV}\,, \label{eq:lepinvmass} \\
  |\eta| &\leq& 2.5\,. \label{eq:CutRap}
\end{eqnarray}
Here $E_{T}(j_1)$ is the transverse energy of the hardest jet, $E_{T}(j)$ is the transverse energy of all other jets, $E_{T}(\ell_{e,\mu})$ is the transverse energy of the leptons, $M_{\ell^+\ell^-}$ is the invariant mass of the opposite-sign same flavour lepton pair, and $\eta$ is the pseudorapidity of all the final state particles in the decay chain. Moreover we require at least two $b$-jets and that 1 $b$-jet plus 2 other jets (typically with the lowest $p_T$) should reconstruct the top quark. Since we need the top momentum to be reconstructed we only take into account tops that decay hadronically.

One of the consequences of the application of the above cuts, especially Eq.~(\ref{eq:leppt}) and (\ref{eq:lepinvmass}) is the significant reduction in the background originating from $\tau$'s, Eq.~(\ref{eq:tautaubackg}). This is due to the rather low energy of the leptons coming from $\tau$ decays and the even lower invariant mass of the resulting lepton pair, which is peaking around 0. Already at this point, approximately 95\% of this background is removed.

Another factor we include is the momenta of the resolved particles being smeared due to the intrinsic experimental precision. The accuracy for both jets and electrons follows the same function but with different coefficients~\cite{:2008zzm}:
\begin{equation}
 \frac{\sigma_{E}}{E} = \sqrt {\frac{a^2_{j,e}}{E} + \frac{b^2_{j,e}}{E^2} + c^2_{j,e}}\ ,
 \label{eq:JetAcc}
\end{equation}
where
 \begin{itemize}
\item for jets $a_j = 0.6\ \mathrm{GeV}^{\frac{1}{2}}$, $b_j = 1.5\ \mathrm{GeV}$ and $c_j = 0.03$;

\item for electrons the accuracy is better, with $a_e = 0.12\ \mathrm{GeV}^{\frac{1}{2}}$, $b_e = 0.2\ \mathrm{GeV}$  and $c_e = 0.01$.
\end{itemize}
The resolution for muons has a different functional dependence,
\begin{equation}
  \frac{\sigma_{p_T}}{p_T} = \Bigg{\{} \begin{array}{cc}  &  0.00008(p_T/\mathrm{GeV}-100)+0.03, \\   & 0.03, \end{array} 
\begin{array}{c} p_T>100 \,\mathrm{GeV}, \\ p_T<100 \,\mathrm{GeV}. \end{array}
\end{equation}
In addition we also have a finite resolution on the measurement of missing transverse energy,
\begin{equation}
 \frac{\sigma^x_{MET}}{E_T}=\frac{\sigma^y_{MET}}{E_T}=\frac{0.57}{\sqrt{E_T/\mathrm{GeV}}}\,,
  \label{eq:METAcc}
\end{equation}
which are the errors on the $x$ and $y$ components of the MET vector and $E_T$ is the scalar sum of all visible transverse energy .

The momenta smearing will only affect the observables when we perform momentum reconstruction as the LSP momenta will be reconstructed with limited precision. The triple product will not suffer however as the measurement only relies on the direction of the measured particles and not on the energy measurement. The direction can be found far more accurately and this error happens to be negligible for our observables.

Finally, we also investigate the fact that the masses of the particles in the decay chains we are interested in will only be known with a certain precision at the LHC (we assume 10\% error), see Section~\ref{sec:scenario}. This error will again only affect the observables when we perform momentum reconstruction in order to boost into the rest frame of the $\tilde{\chi}^0_2$ and can cause the frame to be mis-measured, see Section~\ref{sec:expfactors} for more details.

\subsection{Experimental factors without momentum reconstruction}

\begin{figure}[t!]
\begin{picture}(16,8)
  \put(4.6,7.4){$\mathcal{T}=\overrightarrow{p}_q\cdot(\overrightarrow{p}_{\ell^+}\times\overrightarrow{p}_{\ell^-})$, $\,\,\sqrt{s} = 14\ \mathrm{TeV}$}
 \put(10.7,0.1){$\phi_{1}/\pi$}
 \put(6.2,5.35){\tiny{50 fb$^{-1}$}}
 \put(6.2,4.9){\tiny{100 fb$^{-1}$}}
 \put(6.2,4.4){\tiny{300 fb$^{-1}$}}
 \put(9.3,6.4){3$\sigma$ Observation}
 \put(2.7,7.4){\epsfig{file=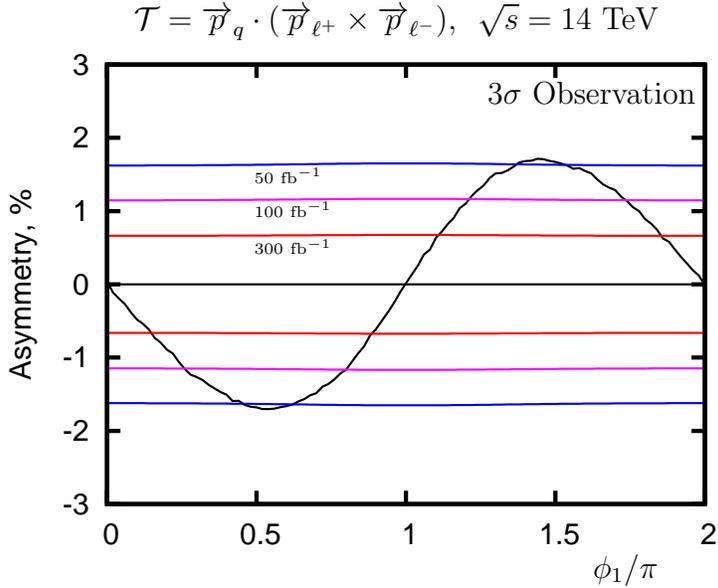,scale=0.8,angle=270}}
\end{picture}
\caption{\label{fig:SqCuts} The asymmetry $\mathcal{A}_{\mathcal{T}}$, Eq.~(\protect\ref{Asy}), in the laboratory frame for the decay chain Eq.~(\protect\ref{eq-decays}) after the cuts, Eqs.~(\protect\ref{eq:CutJH})-(\protect\ref{eq:CutRap}), have been applied. The coloured lines show the size of the asymmetry needed for a $3\sigma$ observation at the given luminosity, $\mathcal{L}$=(50 fb$^{-1}$, 100 fb$^{-1}$, 300 fb$^{-1}$), assuming squarks were produced via the channels shown in rows 3 and 4 of Tab.~\protect\ref{tab:CrossSec}. Momentum smearing for both the leptons and quarks was studied and found to have a negligible effect. All other relevant experimental details are mentioned in the text.}
\vspace{0.cm}
\end{figure}

Out of the experimental factors mentioned above, only the cuts affect the result for the triple product correlation measured in the laboratory frame. These cuts reduce the number of detectable events by $\approx50\%$ and consequently significantly increase the luminosity required to make a statistically significant measurement at the LHC, see Fig.~{\ref{fig:SqCuts}}. For example if large phases are present we may begin to see hints with integrated luminosity $\mathcal{L}=50\ \mathrm{fb}^{-1}$. With $\mathcal{L} = 300\ \mathrm{fb}^{-1}$ we could become sensitive to phases in the ranges $0.15\,\pi \lesssim \phi_{1} \lesssim 0.9\,\pi$ and $1.1\,\pi \lesssim \phi_{1} \lesssim 1.85\,\pi$ where the asymmetry $|\mathcal{A_T}|>0.7$\%.

For the calculation of the asymmetry we included the production channels shown in rows 3 and 4 of Table~\ref{tab:CrossSec} taking into account the decays of individual $\tilde{q}_L$ and $\tilde{q}_L^*$ as listed in Eq~(\ref{eq-decays}). No explicit hadronisation was included but momentum smearing is expected to simulate some of this effect. We also need to include additional hard QCD radiation and other detector effects (for example fakes) in a full experimental study. Correct identification was assumed for the quark and two leptons. The backgrounds in the study are those from $\tilde{q}_L^*$ and Eq.~(\ref{eq:taudec}).

It must be noted however, that significant pollution due to backgrounds will be expected for this signal from both the Standard Model and the MSSM. Further experimental cuts will certainly be required to improve the signal/background ratio and without a more detailed study it is hard to predict what effect this will have on the ability to complete this measurement at the LHC, especially as the asymmetries are rather small. However, a more detailed experimental analysis is beyond the scope of the present paper \cite{newpaper}.

\subsection{Experimental factors with momentum reconstruction}\label{sec:expfactors}

When we perform the momentum reconstruction we need to include the experimental precision on the momentum of the visible particles. This resolution is $\sim3\%$ for leptons and follows Eq.~(\ref{eq:JetAcc}) for jets. The corresponding effect on momentum reconstruction is a reduction in the number of events that have the same sign triple product. As stated in Sec.~\ref{sec:MomRecRes} we discard any events where we have solutions with opposite sign triple products. Discarding these events reduces the percentage we can use from $\sim60\%$ without momentum smearing down to $\sim30\%$ when we include it.

The other difficulty momentum smearing creates is that all the reconstructed solutions can now have the wrong sign triple product as we no longer correctly reproduce the rest frame of the neutralino $\tilde{\chi}^0_2$. Inevitably this effect produces a decrease in the observed asymmetry from $\sim11\%$ to $\sim8\%$.

We again include the cuts on all visible particles in our decay chain given by Eqs.~(\ref{eq:CutJH})-(\ref{eq:CutRap}). These cuts significantly reduce the number of visible events and remove $\sim80\%$ of the events compared with our initial na\"{\i}ve estimates. When we combine the cuts with the momentum reconstruction efficiency we are left with $\sim6\%$ of the initial events and this clearly increases the luminosity needed to make an observation statistically significant. After inclusion of these effects the number of events drops from $20000$ down to $1200$ at the integrated luminosity of $\mathcal{L} = 100\ \mathrm{fb}^{-1}$. This results in a $1\sigma$ absolute uncertainty of order $\sim 3\%$ on the asymmetry, according to Eq.~(\ref{eq:asymmerror}). 

Another possible experimental aspect we investigate is a 10\% uncertainty on the masses of the supersymmetric particles used in the momentum reconstruction. We found that this has a negligible effect on the momentum reconstruction as long as the mass differences between different particles in the decay chain are known better than $\mathcal{O}$(5 GeV). The assumption that the mass differences are known to a higher accuracy is reasonable as the main method of measuring masses in SUSY decay chains at the LHC will be via kinematic end points that are measured with high precision.

\begin{figure}[t!]
\begin{picture}(16,8)
  \put(4.6,7.4){$\mathcal{T}=\overrightarrow{p}_q\cdot(\overrightarrow{p}_{\ell^+}\times\overrightarrow{p}_{\ell^-})$, $\,\,\sqrt{s} = 14\ \mathrm{TeV}$}
 \put(10.7,0.1){$\phi_{1}/\pi$}
 \put(6.2,5.8){\tiny{50 fb$^{-1}$}}
 \put(6.2,5.2){\tiny{100 fb$^{-1}$}}
 \put(6.2,4.60){\tiny{300 fb$^{-1}$}}
 \put(9.3,6.4){3$\sigma$ Observation}
 \put(2.7,7.4){\epsfig{file=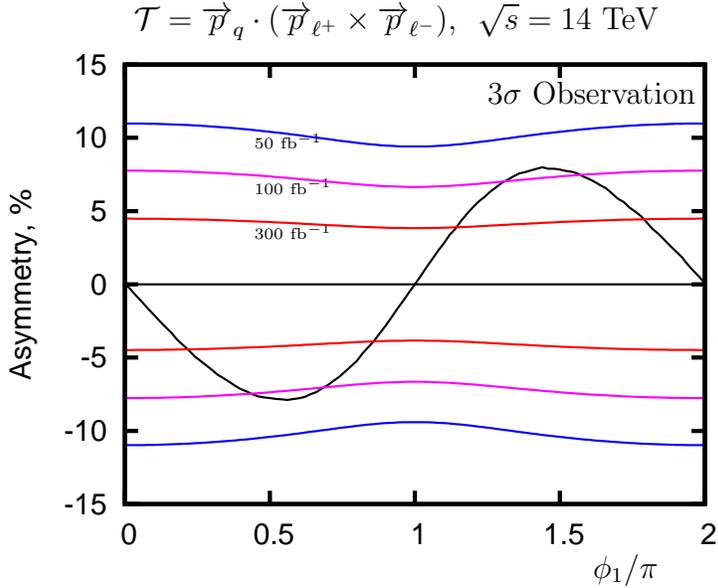,scale=0.8,angle=270}}
\end{picture}
\caption{\label{fig:MomRecCuts} The asymmetry $\mathcal{A}_{\mathcal{T}}$, Eq.~(\protect\ref{Asy}), in the reconstructed $\tilde{q}_L$ frame for $\tilde{q}_L\tilde{g}$, $\tilde{q}_L^*\tilde{g}$ production followed by the decay chains shown in Eq.~(\protect\ref{eq-decays}) and Eq~(\protect\ref{gludec}) with cuts Eqs.~(\protect\ref{eq:CutJH})-(\protect\ref{eq:CutRap}). The coloured lines show the size of the asymmetry needed for a $3\sigma$ observation at the given luminosity, $\mathcal{L}$=(50 fb$^{-1}$, 100 fb$^{-1}$, 300 fb$^{-1}$). The momenta of the final state particles have been smeared according to Eq.~(\protect\ref{eq:JetAcc})-(\protect\ref{eq:METAcc}) to replicate the LHC detector effects. All other relevant experimental details are mentioned in the text.}
\vspace{0.cm}
\end{figure}

Figure~\ref{fig:MomRecCuts} shows the asymmetry and the luminosity required at the LHC to see a statistically significant result at the 3$\sigma$ level at the LHC once all the above factors have been taken into account. The production channels are $\tilde{q}_L\tilde{g}$, $\tilde{q}_L^*\tilde{g}$ and the branching ratios are included from both the $\tilde{q}_L$ and $\tilde{g}$ decay chains. No explicit hadronisation effects were included here and possible additional hard jets in the final state require further study. Also correct identification was assumed for all the leptons and quarks in both decay chains as explained in Sec.~\ref{sec:MomRecRes} and Sec.~\ref{sec:PracApproach}. Again the only backgrounds included in the study are those of $\tilde{q}_L^*\tilde{g}$ production and of taus that decay to visible leptons, Eq.~(\ref{eq:taudec}).

 We can see that using this method, with integrated luminosity $\mathcal{L} \sim 100\, \mathrm{fb}^{-1}$ we start to become sensitive if large phases happen to be present. After $\mathcal{L} \sim 300\, \mathrm{fb}^{-1}$ we expect to have sensitivity to phases in the ranges $0.2\,\pi \lesssim \phi_{1} \lesssim 0.85\,\pi$ and $1.15\,\pi \lesssim \phi_{1} \lesssim 1.8\,\pi$, and obviously more luminosity will improve this further.

Note that a direct comparison between the methods with and without momentum reconstruction based on the above plots should not be performed. The backgrounds from both SM and MSSM will be more severe when we do not perform momentum reconstruction and clearly many new cuts will be required to isolate the signal, which is the consequence of being totally inclusive. For the method when we perform momentum reconstruction, we have a well defined final state that is difficult to be faked by Standard Model processes and is also uncommon for SUSY cascade decays. Moreover the multiple cuts on all the particles and the need for the missing momentum to be successfully reconstructed mean that many backgrounds will be rejected.

\subsection{Ongoing work}
\label{subsec:Ongoing}

A new approach in the future study~\cite{newpaper} is to use matrix element weighting for each of the multiple solutions. This would significantly increase the number of accepted events and possibly increase the statistical significance of the result as we had observed that some of the ``wrong'' solutions obey rather unlikely kinematic configurations. However, the validity of this method needs further study. 

The other possible improvement is the application of the cuts on anti-squarks. In the associated production, Eq.~(\ref{eq-prod}), anti-squarks come always from the sea quarks, whereas squarks can originate both from the sea quarks and the valence quarks. Therefore one can expect some difference in their distributions. This would give a handle to cut out one of our most severe backgrounds. Preliminary studies showed that indeed this selection criteria can turn out to be useful depending on the actual scenario and the collected luminosity.

It is clear that to fully understand if these measurements are possible at the LHC a detector simulation including backgrounds will be required. This study has to include the combinatorial issues that may be encountered with both the final state leptons and the jets. In addition, the impact of particle reconstruction inefficiencies, falsely identified particles (``fakes''), different detector acceptance for different particle species and other experimental considerations need to be considered. Such an analysis is in the line of future work.

 Further improvements in measuring the CP-violating phase $\phi_1$ would require the inclusion of more observables sensitive to the phase, like lepton opening angle or lepton angular distribution \cite{Choi:2005gt}. Nevertheless, in order to determine the phase in a precise and rather model-independent way, experimental input from the clean and unbiased measurements at a future linear collider \cite{Bartl:2004jj,Bartl:2005uh,Bartl:2006yv} is required in addition to the LHC.


\section{Conclusions}

In this paper we have investigated the problem of discovering CP-violating effects at the Large Hadron Collider. Our tool in this analysis was the triple product correlation of momenta of the quark and the lepton pair from the $\tilde{\chi}^0_2$ decay. This kind of T-odd observables can give us a handle to discover the presence of complex phases in the model. We took into account the full production and decay process of squark and gluino. Since triple products depend crucially on spin correlations, they have been included both in the analytical calculation and the event generation, that has been performed using \texttt{Herwig++ 2.3.2}. The process of special interest in our case was the squark decay into quark and neutralino $\tilde{\chi}_2^0$ followed by the three-body leptonic neutralino decay. For this decay in our scenario one can expect an asymmetry in the triple product distribution of up to 14\% when calculated in the rest frame of the decaying neutralino. The source of the CP violation in our case was the phase of the bino mass parameter $M_1$.

Due to the hadronic experimental environment of the LHC, precise measurements will be a challenge both from experimental and theoretical point of view. Therefore we have considered two possible methods of measuring CP-violating effects in the neutralino decay. The first method calculates the triple product in the laboratory frame. It turns out however that the initial CP-odd asymmetry is diluted both by high boosts of produced particles and by the admixture of anti-squarks. This makes its observation far more difficult but can give us hints of CP violation after basic selection cuts in our example.

As mentioned before the largest effects of CP violation can be expected in the rest frame of the decaying neutralino. We studied the use of momentum reconstruction of invisible LSPs and neutrinos to get access to the rest frame of the neutralino $\tilde{\chi}^0_2$. Using a set of invariant kinematic conditions we showed that it is possible to fully reconstruct the production and decay process on an event-by-event basis. The important result here is the possibility of reconstruction in the case of three-body decays and escaping neutrinos.

Having fully reconstructed events we are able to boost particle momenta back to the rest frame of the neutralino $\tilde{\chi}^0_2$ and the asymmetry becomes as large as 10\%. Again, we include experimental event selection criteria as well as momentum smearing to account for finite detector resolution. We also consider the most severe background due to anti-squarks and we showed that the signal contamination from neutralino decays to leptonically decaying tau pairs does not pose a problem for our measurement.

Applying momentum reconstruction and taking into account the experimental effects we have found that the CP-odd asymmetry drops down to 8\%. To avoid ambiguities and due to experimental constraints many events have to be discarded, nevertheless the left-over sample is enough to probe CP-violation effects. One should see a $3\sigma$ effect at $\mathcal{L} = 300\ \mathrm{fb}^{-1}$ and we expect sensitivity to the phase in the range $0.2\,\pi \lesssim \phi_{1} \lesssim 0.85\,\pi$. We emphasise that the asymmetry after momentum reconstruction is a much cleaner observable from a theoretical point of view, thanks to a well defined final state. Therefore, using the above technique there are good prospects for the observation or exclusion of CP-violating effects for a wide range of the phase $\phi_1$ after a few years of LHC running at the high luminosity. The full assessment of LHC's ability to resolve CP violation in the MSSM will definitely require a realistic simulation of detector effects, SM and SUSY backgrounds. That issue is beyond the scope of the present phenomenological analysis and therefore we leave this question for future study. 

The momentum reconstruction procedure turns out to be extremely useful method for studying new physics models at the LHC, as we have shown in this paper. Nevertheless, in order to reveal the underlying model, including CP-violating phases, a more precise machine, like an $e^+e^-$ linear collider, will be required.

\section*{Acknowledgements}
The authors wish to thank Philip Bechtle, Bj\"{o}rn Gosdzik, Bob McElrath and Filip Moortgat for valuable discussions. We also are grateful to David Grellscheid and Peter Richardson for their assistance in the use of \texttt{Herwig++}.
KR is supported by the EU Network MRTN-CT-2006-035505 ``Tools and Precision Calculations for Physics Discoveries at Colliders'' (HEPTools). JT is supported by the UK Science and Technology Facilities Council (STFC).

\newpage
\section*{Appendices}

\begin{appendix}

\section{Interaction Lagrangian and neutralino mixing}
\label{sec:lagrangian-couplings}

Neglecting the mixing effects in the slepton sector, the interaction Lagrangian terms for $\tilde{\chi}^0_2$ decay are
\begin{eqnarray}
 \label{eq:LagZll}
 && \mathcal{L}_{Z^0\ell^+\ell^-} =
  -\frac{g}{\cos\theta_W}Z_{\mu}\bar{\ell}\gamma^{\mu}
  [L_\ell P_L + R_\ell P_R]\ell\,, \\
&& \mathcal{L}_{Z^0\tilde{\chi}^0_m\tilde{\chi}^0_n} =
 \frac{g}{2 \cos\theta_W} Z_{\mu}{\bar{\tilde{\chi}}}^0_m\gamma^{\mu}
 [{O}_{mn}^L P_L + {O}_{mn}^R P_R]{\tilde{\chi}}^0_n\,, \\
 && \mathcal{L}_{\ell\tilde{\ell}\tilde{\chi}^0_k} =
 {g} f_{Lk}^{\ell}\: \bar{\ell}\: P_R\: \tilde{\chi}^0_k\: \tilde{\ell}_L + {g} f_{Rk}^{\ell}\: \bar{\ell}\: P_L\: \tilde{\chi}^0_k\: \tilde{\ell}_R +
 \textrm{h.c.}\,,
\end{eqnarray}
where $P_{L,R}=\frac{1}{2}(1\mp\gamma_5)$, $g = e/\sin\theta_W$ and $\theta_W$ is the
weak mixing angle. The couplings are given by
 \begin{eqnarray}
  f_{Lk}^{\ell} &=& \frac{1}{\sqrt{2}} \, (\tan\t_W N_{k1} +  {N_{k2}})\,,  \\
  f_{Rk}^{\ell} &=& -\sqrt{2}\, \tan\t_W  N_{k1}^* \,, \\
{O}_{mn}^L &=& N_{m4} N_{n4}^* -  N_{m3} N_{n3}^*\,,\\
{O}_{mn}^R &=& - {O}_{mn}^{L*}\,,\\
L_\ell &=& -\frac{1}{2} + \sin^2\theta_W\,,\\
R_\ell &=& \sin^2\theta_W\,,
\end{eqnarray}
where $N$ is the neutralino mixing matrix defined by Eq.~(\ref{eq:Nmatrix}).

\vspace{0.5cm}

The interaction Lagrangian for $\tilde{q}_L$ decay is
\begin{equation}
\mathcal{L}_{q\sq \tilde{\chi}^0_k} =
  g f^{q}_{Lk}\: \bar{q}\: P_R\: \tilde{\chi}^0_k\: \sq_L+ \mathrm{h.c.}, 
\end{equation}
with
 \begin{eqnarray}
  f_{Lk}^{q} &=&  \frac{1}{3 \sqrt{2}} \, (- T_{3q} \tan\t_W N_{k1} + 3  {N_{k2}})\,, \label{eq:fLk}
 \end{eqnarray}
where $T_{3q}$ is a weak isospin of a squark and we have neglected small Yukawa couplings for the first two generations of squarks and mixing between left and right squark states.

\vspace{0.5cm}

The complex symmetric neutralino mass matrix ${\cal M}_N$ is given in the basis
$(\tilde{B}, \tilde{W}^3, \tilde{H}^0_1, \tilde{H}^0_2)$ by:
\begin{equation}
  {\cal M}_N = \left(
    \begin{array}{cccc}
      M_1  &  0  & -m_Z c_\beta s_W & m_Z s_\beta s_W \\
      0 & M_2 & m_Z c_\beta c_W & -m_Z s_\beta c_W \\
      -m_Z c_\beta s_W  & m_Z c_\beta c_W  & 0 & -\mu  \\
      m_Z s_\beta s_W & -m_Z s_\beta c_W & -\mu & 0 \\
    \end{array} \right)\,,
\end{equation}
where the abbreviations $s_W=\sin\theta_W,\, c_W=\cos\theta_W,\, s_{\beta}=\sin\beta,\, c_{\beta}=\cos\beta$
have been used and $\tan\beta=v_2/v_1$ is the ratio of the vacuum expectation values of the Higgs fields. It is diagonalised by the unitary $4\times4$ matrix $N$ 
\begin{equation} 
{\rm diag}(m_{\tilde{\chi}_1^0},m_{\tilde{\chi}_2^0},m_{\tilde{\chi}_3^0},m_{\tilde{\chi}_4^0}) = N^*{\cal M}_N N^\dagger\,,\,\,\,\,\,\,\,\,\,\,
(m_{\tilde{\chi}_1^0}<m_{\tilde{\chi}_2^0}<m_{\tilde{\chi}_3^0}<m_{\tilde{\chi}_4^0})\label{eq:Nmatrix}
\end{equation}
giving 4 non-negative Majorana neutralino mass eigenstates. The masses in Eq.~(\ref{eq:Nmatrix}) can be chosen real and positive by a suitable definition of the unitary matrix $N$. For convenience, in the following we use the notation: $m_i \equiv m_{\tilde{\chi}_i^0}$.

\section{Amplitude squared including full spin correlations}
\subsection{Neutralino production $\tilde{q}_L \to \tilde{\chi}^0_j q$\label{sect:squarkdecay}}

The analytic expression for the production density matrix can be decomposed as
\begin{equation}
|M(\tilde{q}_L\to \tilde{\chi}^0_j q)|^2=P(\tilde{\chi}^0_j q)+\Sigma^a_P(\tilde{\chi}^0_j)\,,
\label{eq_mq}
\end{equation}
whose spin-independent contribution reads
\begin{equation}
P(\tilde{\chi}^0_j q)=\frac{g^2}{2} |f_{Lj}^{q}|^2 (p_q p_{\tilde{\chi}^0_j})\,,
\end{equation}
where $p_q$ and $p_{\tilde{\chi}^0_j}$ denote the four-momenta of the quark $q$ and 
the neutralino $\tilde{\chi}^0_j$. The spin-dependent contributions is T-even and given by
\begin{equation}
\Sigma^a_P(\tilde{\chi}^0_j) = - \frac{g^2}{2} |f_{Lj}^{q}|^2 m_j 
(p_q s^a(\tilde{\chi}^0_j)) \,,\label{eq_prod-ea}
\end{equation}
where $s^a(\tilde{\chi}^0_j)$ denotes the spin-basis vector of the neutralino $\tilde{\chi}^0_j$. 
The three spin-basis four-vectors $s^1$, $s^2$ and $s^3$ form a right-handed system and provide, together with the momentum, an orthogonal basis system.  They are chosen as:
\begin{eqnarray}
s^1(\tilde{\chi}^0_j)&=&\left(0,\frac{(\vec{p}_{\tilde{\chi}^0_j}\times \vec{p}_{\tilde{q}_L})\times \vec{p}_{\tilde{\chi}^0_j}}{|(\vec{p}_{\tilde{\chi}^0_j}\times \vec{p}_{\tilde{q}_L})\times \vec{p}_{\tilde{\chi}^0_j}|}\right)\,, \label{eq-s1chi}\\
s^2(\tilde{\chi}^0_j)&=&\left(0,\frac{\vec{p}_{\tilde{\chi}^0_j}\times \vec{p}_{\tilde{q}_L }} {|\vec{p}_{\tilde{\chi}^0_j}\times \vec{p}_{\tilde{q}_L}|}\right)\,,
\label{eq-s2chi}\\
s^3(\tilde{\chi}^0_j)&=&\frac{1}{m_j}\left( |\vec{p}_{\tilde{\chi}^0_j}|, \frac{E_{\tilde{\chi}^0_j}}{|\vec{p}_{\tilde{\chi}^0_j}|} \vec{p}_{\tilde{\chi}^0_j}\right)\,.
\label{eq-s3chi}
\end{eqnarray}
For a more detailed discussion see \cite{Ellis:2008hq}.

\subsection{Neutralino three-body decay $\tilde{\chi}^0_j \to
\tilde{\chi}^0_k \ell^+ \ell^-$ \label{sect:neutdecay}}
Here we give the analytical expressions for the
different contributions to the
decay density matrix
for the three-body leptonic neutralino decay, where we sum over
the spins of the final-state particles~\cite{MoortgatPick:1999di}.
The contributions independent of the polarisation of the neutralino
$\tilde{\chi}^0_j$
\begin{equation}
D(\tilde{\chi}^0_j)=D(Z Z)+ D(Z \tilde{\ell}_L)+ D(Z
\tilde{\ell}_R)+ D(\tilde{\ell}_L \tilde{\ell}_L)+
D(\tilde{\ell}_R \tilde{\ell}_R) \label{eq_sumz}\,,
\end{equation}
are given by
\begin{eqnarray} D(Z Z)&=& 8
  \frac{g^4}{\cos^4\theta_W} |\Delta(Z)|^2
  (L_{\ell}^2+R_{\ell}^2) \nonumber\\
 & & \times \Big[ |O^{L}_{kj}|^2 (g_1+g_2) +(\mathrm{Re}\:
  O^{L}_{kj})^2 -(\mathrm{Im}\: O^{L}_{kj})^2) g_3  \Big]\,,
  \label{eq_dzz}\\
D(Z \tilde{\ell}_L)&=&4 \frac{g^4}{\cos^2\theta_W} L_{\ell}\,
  \mathrm{Re}\:\Big\{\Delta(Z) \Big[f^{\ell}_{L j} f^{\ell*}_{L k}
  \Delta_t^{*}(\tilde{\ell}_L)
  (2O^{L}_{kj} g_1 +O^{L*}_{kj} g_3) \nonumber\\
 & &\phantom{4 \frac{g^4}{\cos^2\theta_W} L_{\ell}
     Re\Big\{\Delta(Z) \Big[}  
  +f^{\ell*}_{L j} f^{\ell}_{L k} \Delta_u^{*}(\tilde{\ell}_L)
  (2O^{L*}_{kj} g_2 +O^{L}_{kj} g_3)
  \Big]\Big\}\,,\label{eq_dzel}\\
D(\tilde{\ell}_L \tilde{\ell}_L)&=& 2 g^4 \Big[ |f^{\ell}_{L j}|^2
  |f^{\ell}_{L k}|^2 \big(|\Delta_t(\tilde{\ell}_L)|^2 g_1
  +|\Delta_{u}(\tilde{\ell}_L)|^2 g_2\big)\nonumber\\
 & &\phantom{2 g^4 \Big[} +\mathrm{Re}\:\big\{(f^{\ell*}_{L j})^2 (f^{\ell}_{L k})^2
  \Delta_{t}(\tilde{\ell}_L) \Delta_u^{*}(\tilde{\ell}_L)\big\}
  g_3 \Big]\,,  \label{eq_delel}
\end{eqnarray}
where $\Delta(Z)$ and $\Delta_{t,u}(\tilde{\ell}_{L})$ denote the propagators of the 
virtual particles in the direct channel and in both crossed channels (labelled $t,u$, cf.\ 
Fig.~\ref{Fig:FeynDecayA}).
The quantities $D(Z\tilde{\ell}_R)$ and $D(\tilde{\ell}_R \tilde{\ell}_R)$
can be derived from Eqs.~(\ref{eq_dzel}) and (\ref{eq_delel}) by
the substitutions
\begin{equation} \label{eq_substdecayP}
 L_{\ell}\to R_{\ell}\,, \quad
 \Delta_{t,u}(\tilde{\ell}_L)\to \Delta_{t,u}(\tilde{\ell}_R)\,,\quad
 O^{L}_{kj}\to O^{R}_{kj}\,, \quad
 f_{L j,k}^{\ell}\to f_{R j,k}^{\ell}\,.
\end{equation}
The kinematic factors are
\begin{eqnarray}
g_1&=&(p_{\tilde{\chi}^0_k} p_{\ell^-})(p_{\tilde{\chi}^0_j} p_{\ell^+})\,,
\label{eq_dkin1}\\
g_2&=& (p_{\tilde{\chi}^0_k} p_{\ell^+})(p_{\tilde{\chi}^0_j} p_{\ell^-})\,,
\label{eq_dkin2}\\
g_3 &=& m_j m_k (p_{\ell^-} p_{\ell^+})\,.\label{eq_dkin3}
\end{eqnarray}
We can split the terms depending on the 
polarisation of the neutralino into T-even and T-odd 
contributions:
\begin{equation}
\Sigma_D^{a}(\tilde{\chi}^0_j)= \Sigma_D^{a,E}(\tilde{\chi}^0_j)
+\Sigma_D^{a,O}(\tilde{\chi}^0_j)\,.
\end{equation}
The T-even contributions depending on the polarisation of the decaying
neutralino $\tilde{\chi}^0_j$
\begin{equation}
\Sigma_D^{a,E}(\tilde{\chi}^0_j)= \Sigma_D^{a,E}(ZZ) +\Sigma_D^{a,E}(Z
\tilde{\ell}_L) +\Sigma_D^{a,E}(Z \tilde{\ell}_R)
+\Sigma_D^{a,E}(\tilde{\ell}_L \tilde{\ell}_L)
+\Sigma_D^{a,E}(\tilde{\ell}_R \tilde{\ell}_R)\, \label{eq_dssum-e}
\end{equation}
are
\begin{eqnarray}
\Sigma_D^{a,E}(ZZ)&=& 8 \frac{g^4}{\cos^4\theta_W} |\Delta(Z)|^2
  (R^2_{\ell}-L_{\ell}^2) \nonumber\\
 & &\times\Big[ -[(\mathrm{Re}\: O^{L}_{kj})^2 -(\mathrm{Im}\: O^{L}_{kj})^2]g^a_3+
  |O^{L}_{kj}|^2(g^a_1-g^a_2) \Big]\,,\label{eq_dszz}\\
\Sigma_D^{a,E}(Z \tilde{\ell}_L)&=& \frac{4
  g^4}{\cos^2\theta_W}L_{\ell}\, \mathrm{Re}\:\Big\{\Delta(Z)
  \Big[ f^{\ell}_{L j} f^{\ell*}_{L k} \Delta_t^{*}(\tilde{\ell}_L)
  \big(-2 O^{L}_{kj} g^a_1 +O^{L*}_{kj} g^a_3\big)\nonumber\\
 & & \phantom{\frac{4 g^4}{\cos^2\theta_W}L_{\ell}\, Re\Big\{}
  + f^{\ell*}_{L j} f^{\ell}_{L k} \Delta_u^{*}(\tilde{\ell}_L)
  \big(2 O^{L*}_{kj} g^a_2 +O^{L}_{kj} g^a_3\big)
  \Big]\Big\}\,,\label{eq_dszel}\\
\Sigma_D^{a,E}(\tilde{\ell}_L \tilde{\ell}_L) & = & 2 g^4 \Big[
  |f^{\ell}_{L j}|^2 |f^{\ell}_{L k}|^2
  [|\Delta_{u}(\tilde{\ell}_L)|^2 g_2^a
  -|\Delta_{t}(\tilde{\ell}_L)|^2 g_1^a]\nonumber\\
& & \phantom{2 g^4 \Big[}
  + \mathrm{Re}\:\big\{ (f^{\ell*}_{L j})^2 (f^{\ell}_{L k})^2
  \Delta_{t}(\tilde{\ell}_L)
  \Delta_u^{*}(\tilde{\ell}_L)g_3^a\big\}\Big]\,,
  \label{eq_dselel}
\end{eqnarray}
where the contributions $\Sigma^{a,E}_D(Z\tilde{\ell}_R)$ and
$\Sigma^{a,E}_D(\tilde{\ell}_R \tilde{\ell}_R)$ are derived from
Eqs.~(\ref{eq_dszel}) and (\ref{eq_dselel}) by applying the substitutions
given by Eq.~(\ref{eq_substdecayP}) and in addition $g_{1,2,3}^a\to -g_{1,2,3}^a$.
The kinematic factors are
\begin{eqnarray}
g^a_1&=& m_j (p_{\tilde{\chi}^0_k} p_{\ell^-}) (p_{\ell^+} 
s^a)\,, \label{eq422_3a}\\
g^a_2&=& m_j (p_{\tilde{\chi}^0_k} p_{\ell^+}) 
(p_{\ell^-} s^a)\,, \label{eq_dssub4}\\
g^a_3&=& m_k [(p_{\tilde{\chi}^0_j} p_{\ell^+}) (p_{\ell^-} s^a)
-(p_{\tilde{\chi}^0_j} p_{\ell^-}) (p_{\ell^+} s^a)]\,.
\label{eq_dssub5}
\end{eqnarray}

The T-odd contributions depending on the polarisation of the decaying
neutralino $\tilde{\chi}^0_j$
\begin{equation}
\Sigma_D^{a,O}(\tilde{\chi}^0_j)= \Sigma_D^{a,O}(ZZ) +\Sigma_D^{a,O}(Z
\tilde{\ell}_L) +\Sigma_D^{a,O}(Z \tilde{\ell}_R)
+\Sigma_D^{a,O}(\tilde{\ell}_L \tilde{\ell}_L)
+\Sigma_D^{a,O}(\tilde{\ell}_R \tilde{\ell}_R)\, \label{eq_dssum-o}
\end{equation}
are
\begin{eqnarray}
\Sigma_D^{a, \mathrm{O}}(ZZ) &=& 
    8 \frac{g^4}{\cos^4\theta_W} |\Delta(Z)|^2
    (L^2_{\ell} - R_{\ell}^2) 
    \Big[ 2 \mathrm{Re}\:(O^{L}_{kj})\, \mathrm{Im}\:(O^{L}_{kj}) i g_4^a \Big]\,,
    \label{eq_dszz_to}\\
\Sigma_D^{a, \mathrm{O}}(Z \tilde{\ell}_L) & = & 
    \frac{4 g^4}{\cos^2\theta_W}L_{\ell}\,
    \mathrm{Re}\:\Big\{\Delta(Z)
    \Big[-f^{\ell}_{L j} f^{\ell*}_{L k} O^{L*}_{kj}
    \Delta_t^{*}(\tilde{\ell}_L)\nonumber\\
  & &\phantom{\frac{4 g^4}{\cos^2\Theta_W}L_{\ell}Re\Big\{\Delta(Z)\Big[}
    + f^{\ell*}_{L j} f^{\ell}_{L k} O^{L}_{kj}
    \Delta_u^{*}(\tilde{\ell}_L)
    \Big]g_4^a\Big\}\,,\label{eq_dszel_to}\\
\Sigma_D^{a, \mathrm{O}}(\tilde{\ell}_L \tilde{\ell}_L)&=& 
    2 g^4\, \mathrm{Re}\:\Big\{ (f^{\ell*}_{L j})^2 (f^{\ell}_{L k})^2
    \Delta_{t}(\tilde{\ell}_L)
    \Delta_u^{*}(\tilde{\ell}_L) g_4^a\Big\}\,,
    \label{eq_dselel_to}
\end{eqnarray}
where the contributions $\Sigma^{a,O}_D(Z\tilde{\ell}_R)$ and
$\Sigma^{a,O}_D(\tilde{\ell}_R \tilde{\ell}_R)$ are derived from
Eqs.~(\ref{eq_dszel}) and (\ref{eq_dselel}) by applying the substitutions
given by Eq.~(\ref{eq_substdecayP}).
The kinematic factor is
\begin{eqnarray}
g^a_4 & = & i m_k \epsilon_{\mu \nu \rho \sigma}
 s^{a \mu} p_{\tilde{\chi}^0_j}^{\nu} p_{\ell^-}^{\rho} p_{\ell^+}^{\sigma}\,. 
\label{eq_dssub6}
\end{eqnarray}

\section{Kinematics}
\label{sec:kinematics}
  \subsection{Phase space}
  The complete cross section for the process can be decomposed into the production 
cross section and the branching ratios of the subsequent decays:
  \begin{eqnarray}
     d\sigma_{\mathrm{tot}} &=& d{\sigma}(qg \to \sq_L \tilde{g})\cdot\frac{E_{\sq_L
}}{m_{\sq_L}\Gamma_{\sq_L}}\,
     d\Gamma  (\sq_L \to q \tilde\chi^0_2)\,\cdot \, \frac{E_{\tilde{\chi}^0_2}}{m_{2} \Gamma_{\tilde{\chi}^0_2}}
        \,d\Gamma (\tilde\chi^0_2\to \tilde{\chi}^0_1 \ell^+ \ell^-)\, ,\label{dGamma}
      \end{eqnarray}
where the factors $E/m \Gamma$ come from the use of the narrow-width approximation 
for the propagators of the $\sq_L$ and $\tilde{\chi}^0_2$. This approximation is valid for 
$(\Gamma/m)^2 \ll 1$, which is satisfied for $\Gamma_{\sq_L} \sim 4.7$~GeV and for $\Gamma_{\tilde{\chi}^0_2}\sim 3\times10^{-5}$~GeV where the width is small because only the three-body decay is kinematically possible. 

We have:
     \begin{eqnarray}
          d\Gamma  (\sq_L \to q  \tilde\chi^0_2)
                 &=&\frac2{E_{\sq_L}}P(\tilde{\chi}^0_2q) \,d\Phi_{\sq}\,, \\
            d\Gamma  (\tilde{\chi}^0_2 \to \tilde{\chi}^0_1 \ell^+ \ell^-)
                 &=&\frac1{4 E_{\tilde{\chi}^0_2}}D(\tilde{\chi}^0_2)\,d\Phi_{\tilde{\chi}^0_2}\,,
     \end{eqnarray}
where the phase-space factors in the laboratory system are given by:
      \begin{eqnarray}
          d\Phi_{\sq}
                 &=& \frac{1}{(2\pi)^2} \frac{E_q}
        {2||\vec{p}_{\tilde{q}}|\cos\theta_{\tilde{q}}-E_{\tilde{\chi}^0_2}-E_q|}~d\Omega_{\sq}\,, \\
            d\Phi_{\tilde{\chi}^0_2 }
                 &=&\frac1{8(2\pi)^5} \frac{E_{\ell^+}}{||\vec{p}_{\tilde{\chi}^0_2}|
\cos \theta_{\ell^+}-E_{\tilde{\chi}^0_1}-E_{\ell^+}-E_{\ell^-} \cos\alpha|}~E_{\ell^-}
dE_{\ell^-}d\Omega_{\ell^+}d\Omega_{\ell^-}\,,
     \end{eqnarray}
where $\alpha$ is the opening angle between the two final state leptons.
\subsection{Integration limits}
     When evaluating the phase-space integral at the parton level, kinematical limits
     need to be determined on the integration variables and these are listed below.
The three-body decay phase space of the $\tilde{\chi}^0_2$ has the following limits:
\begin{eqnarray}
       E_{\ell^-}&<&\frac{m_2^2-m_1^2}{2(E_{\tilde{\chi}^0_2}-|\vec{p}_{\tilde{\chi}^0_2}|)}\,, \\
       \cos\theta_{\ell^-}&<&\frac{2E_{\tilde{\chi}^0_2}E_{\ell^-}+m_1^2-m_2^2}{2E_{\ell^-}|\vec{p}_{\tilde{\chi}^0_2}|}\,.
\end{eqnarray}
For a more detailed discussion see \cite{Ellis:2008hq}.
\end{appendix}


\addcontentsline{toc}{section}{References}
\bibliography{refs}

\end{document}